\documentclass[11pt,a4paper]{article}
\pdfoutput=1
\usepackage{jcappub}
\usepackage{rotating}
\usepackage{array}
\usepackage{amsmath}
\usepackage[normalem]{ulem}
\usepackage{cancel}
\usepackage{slashed}
\usepackage{booktabs}
\usepackage[pdftex,table]{xcolor}
\usepackage{units}
\usepackage{xspace}
\usepackage{subcaption}
\usepackage{graphicx}
\usepackage{algorithm}
\usepackage{algpseudocode}
\usepackage{rotating}
\usepackage{adjustbox}
\usepackage{tabularx}
\usepackage{bm}
\usepackage{rotating}
\usepackage{caption}

\newcommand{\ba}[1]{\ensuremath{\left( #1 \right)}}
\newcommand{\bb}[1]{\ensuremath{\left[ #1 \right]}}
\newcommand{\bc}[1]{\ensuremath{\left\{ #1 \right\}}}

\newcommand*\diff{\mathop{}\!\mathrm{d}}

\newcommand{\vw}{\ensuremath{v_\text{w}}}

\newcommand{\DS}{\ensuremath{\text{DS}}}
\newcommand{\SM}{\ensuremath{\text{SM}}}
\newcommand{\tot}{\ensuremath{\text{tot}}}
\newcommand{\DNeff}{\ensuremath{\Delta N_\text{eff} }}
\newcommand{\perc}{\ensuremath{\text{perc}}}
\newcommand{\sw}{\ensuremath{\text{sw}}}
\newcommand{\bw}{\ensuremath{\text{bw}}}
\newcommand{\reh}{\ensuremath{\text{reh}}}
\newcommand{\obs}{\ensuremath{\text{obs}}}
\newcommand{\cd}{\ensuremath{\text{cd}}}

\definecolor{DESYcyan}{RGB}{0,159,223}
\definecolor{DESYorange}{RGB}{241,143,31}
\definecolor{DESYrot}{RGB}{235,90,45}
\definecolor{DESYdunkelrot}{RGB}{185,45,65}
\definecolor{DESYdunkelblau}{RGB}{0,75,110}
\definecolor{DESYviolett}{RGB}{145,125,185}
\definecolor{DESYlila}{RGB}{80,80,155}
\definecolor{DESYgelb}{RGB}{250,200,0}

\definecolor{PlotGreen}{HTML}{3cb44b}
\definecolor{PlotRed}{HTML}{e6194b}
\definecolor{PlotYellow}{RGB}{255, 225, 25}

\usepackage{multirow}

\arxivnumber{DESY-23-077}

\title{Does NANOGrav observe a dark sector phase transition?}

\author[1]{Torsten Bringmann,}
\author[2]{Paul Frederik Depta,}
\author[3]{Thomas Konstandin,}
\author[3]{Kai Schmidt-Hoberg,}
\author[3]{and Carlo Tasillo}

\affiliation[1]{Department of Physics, University of Oslo, Box 1048, N-0316 Oslo, Norway}
\affiliation[2]{Max-Planck-Institut f\"ur Kernphysik, Saupfercheckweg 1, 69117 Heidelberg, Germany}
\affiliation[3]{Deutsches Elektronen-Synchrotron DESY, Notkestr.~85, 22607 Hamburg, Germany}

\emailAdd{torsten.bringmann@fys.uio.no}
\emailAdd{frederik.depta@mpi-hd.mpg.de}
\emailAdd{thomas.konstandin@desy.de}
\emailAdd{kai.schmidt-hoberg@desy.de}
\emailAdd{carlo.tasillo@desy.de}

\abstract{Gravitational waves from a first-order cosmological phase transition, at temperatures at the MeV-scale, would 
arguably be the most exciting explanation of the common red spectrum reported by the NANOGrav collaboration, not the least 
because this would be direct evidence of physics beyond the standard model. Here we perform a detailed analysis of whether 
such an interpretation is consistent with constraints on the released energy deriving from big bang nucleosynthesis and the 
cosmic microwave 
background. We find that a phase transition in a completely secluded dark sector is strongly disfavoured with respect to 
the more conventional astrophysical explanation of the putative gravitational wave signal in terms of supermassive black hole 
binaries.  On the other hand,
a phase transition in a dark sector that subsequently decays, before the time of 
neutrino decoupling, remains an intriguing possibility to explain the data. 
From the model-building perspective, such an option is easily satisfied for couplings with the visible sector that are 
small enough to evade current collider and astrophysical constraints. 
The first indication that could eventually corroborate such an interpretation, once the observed common red spectrum 
is confirmed as a nHz gravitational wave background,
could be the spectral tilt 
of the signal. In fact, the current data already show a very slight preference for a spectrum that is softer than 
what is expected from the leading astrophysical explanation.
}
\keywords{big bang nucleosynthesis, cosmological phase transitions, particle physics --- cosmology connection}

\begin{document}
\maketitle
\flushbottom

\section{Introduction}

Gravitational wave (GW) observatories provide us with an additional pair of eyes --- or rather ears --- to explore the 
Universe~\cite{LIGOScientific:2007fwp, LISA:2017pwj, NANOGrav:2020bcs}. Since the first detection of GWs in 2015~\cite{LIGOScientific:2016aoc} they help us to better understand our cosmic 
neighbourhood. With more observatories joining in international collaborations and new 
experiments being consistently proposed, we can eventually 
hope to not only detect local, astrophysical sources of gravitational radiation, but also GWs of cosmological origin. 
To provide context, observables related to primordial nucleosynthesis currently constitute the earliest robust test of the 
cosmological concordance 
model~\cite{Burles:2000zk}. GWs could in principle be used to directly probe the cosmological history at even earlier
times, because they propagate essentially unperturbed after their production.
The corresponding cosmological GW signals will generally form a stochastic gravitational wave background (GWB), in many 
ways similar to the cosmic microwave background (CMB)~\cite{Auclair2022}. 

In 2020 the North American Nanohertz Observatory for Gravitational Waves (NANO\-Grav) collaboration announced 
strong evidence for a so-called common red process in the nHz frequency range~\cite{NANOGrav:2020bcs}, 
i.e.~a stochastic signal that features the same spectrum of temporal correlations
in the pulse arrival times among the set of analysed pulsars.
To verify that the signal is really the first detection of a stochastic GWB, the characteristic quadrupolar 
``Hellings-Downs'' (HD) inter-pulsar correlation~\cite{Hellings1983} will have to be confirmed. 
While this correlation is not expected to be seen in the current data set in view of the still relatively low signal-to-noise 
ratios~\cite{Romano2020}, its confirmation may be imminent with the upcoming data release.
Following NANOGrav's announcement, also the Parkes Pulsar Timing Array (PPTA)~\cite{Goncharov2021}, the European Pulsar Timing Array (EPTA)~\cite{Chen2021} as well as the International Pulsar Timing Array (IPTA)~\cite{Antoniadis2022} published comparable hints for a GWB (see however Zic et al.~\cite{Zic2022}). 
Follow-up data releases including more correlation data are thus both anticipated and eagerly awaited, 
and will contribute to a better understanding of the origin of the observed common red signal~\cite{Pol2020}.

Assuming a confirmation of the inter-pulsar correlations as a GWB, 
it will be fascinating to further study its properties and to determine the underlying physical processes that generated the GWs.
Expected astrophysical sources in this frequency range are mergers of supermassive black hole binaries 
(SMBHB)~\cite{NANOGrav:2020bcs}. In order to explain the observed signal amplitude, however,
the local SMBHB density would need to be  higher by a factor of a few compared to previous 
estimates~\cite{Casey-Clyde2021,Kelley2016, Kelley2017},  
and the question of whether realistic astrophysical models could give rise to a sufficiently strong GWB signal remain 
the subject of an ongoing debate~\cite{Middleton2021,Izquierdo-Villalba2021,Curylo2021,Somalwar2023}.
An alternative and arguably even more exciting possibility is that the signal could be of truly cosmological origin, 
i.e.~the GWB could have been generated by physical processes at redshifts much larger than the $z\lesssim2.5$ associated with 
the SMBHB hypothesis. Cosmological 
sources that have been investigated in the context of the tentative NANOGrav signal include GW emission due to inflation~\cite{Vagnozzi:2020gtf}, first-order phase 
transitions~\cite{Nakai2020,Ratzinger:2020koh, NANOGrav:2021flc}, cosmic strings~\cite{Blasi2020,Ellis2020b, Buchmuller:2020lbh} or the formation of primordial black 
holes~\cite{Luca2020, Kohri2020}. 

In this work we examine how and under which circumstances first-order cosmological phase transitions 
can provide a viable explanation of the signal. Such a phase transition is not present in the 
standard model (SM) of particle physics~\cite{Kajantie:1996mn}, but there are many well-motivated options for how it
could be induced by symmetry breaking or confinement connected to physics beyond the SM~\cite{Caprini2015}. 
In order to match the putative GW signal at nHz frequencies,
the preferred temperature for such a phase transition must be at the MeV-scale~\cite{NANOGrav:2021flc}. As new physics at this 
energy scale is very strongly constrained by a large variety of direct experimental searches~\cite{Gori:2022vri}, 
this immediately implies that the associated new states should only couple very weakly to the SM. 
In other words, such a phase transition would have to take place in a more or less secluded
dark sector (DS) --- which could, in fact, also directly be related to the dark matter 
puzzle~\cite{Pospelov:2007mp,Feng:2008mu,Pospelov:2008zw}.
Importantly, even if the DS is only very weakly coupled to the SM, a DS phase transition (DSPT) could impact the 
successful predictions of Big Bang Nucleosynthesis (BBN) and the CMB due to the extra energy density that is present in the DS
(conventionally parametrised as an effective number of new relativistic neutrino degrees of freedom, $\Delta N_\text{eff}$)~\cite{Planck2018,Yeh2022} or through the late decay of the additional states~\cite{Hufnagel:2018bjp,Forestell:2018txr,Depta2020,Depta:2020mhj,Kawasaki:2020qxm}.
In this work we will consider the two main generic possibilities, namely where the additional energy density
\begin{enumerate}
\item  fully remains within the DS (``stable DS''), or 
\item is subsequently injected into the SM sector (``decaying DS'').
\end{enumerate}
It is worth stressing that these options simply refer to different regimes of the inter-sector coupling(s), and hence
are {\it a priori} equally viable from a phenomenological point of view.
In both cases, in particular, we do not assume those inter-sector couplings to be sufficiently large for the DS to thermalise with the 
SM heat bath. As a result, for strong enough couplings within the DS, the visible and dark sector will generally have different 
temperatures~\cite{Breitbach2019}.

\begin{figure}[t]
	\centering
	\includegraphics[width=0.85\linewidth]{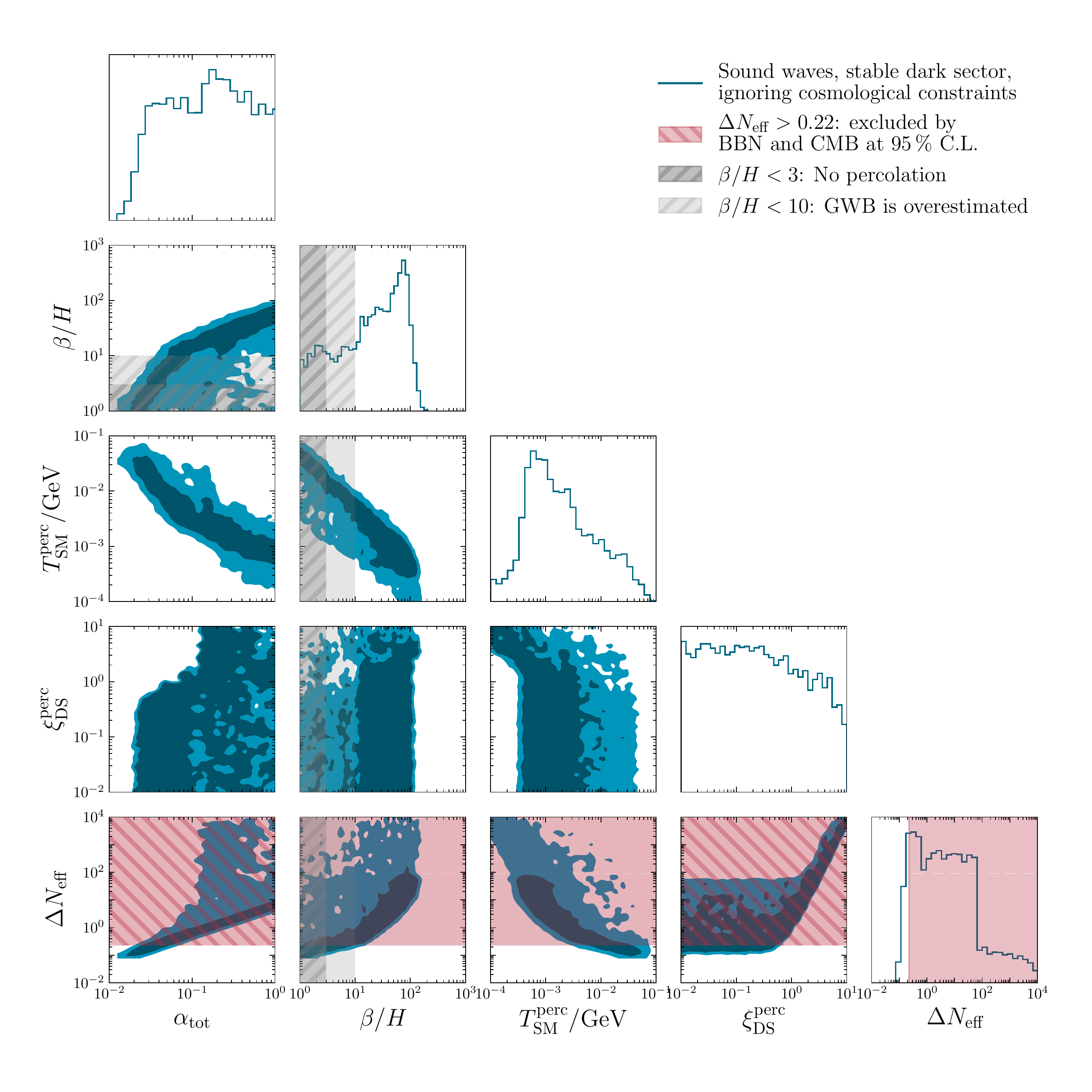}
	\caption{Triangle plot showing the $1\,\sigma$ and $2\,\sigma$ contours
	obtained by a naive fit (\textcolor{DESYdunkelblau}{blue}) 
	of the NANOGrav 12.5 year data to a GW spectrum emitted in a DS phase transition, ignoring cosmological constraints. 
	To illustrate the tension with 
	BBN and CMB, the 95\% C.L. excluded regions corresponding to $\Delta N_\text{eff} > 0.22$ are shaded in 
	\textcolor{DESYdunkelrot}{red}~\cite{Yeh2022}, cf.~the discussion in section~\ref{sec:constraints}. The regions in which 
	no percolation ($\beta/H < 3$) and an overestimation of the GWB amplitude ($\beta/H < 10$) are expected are shaded in \textcolor{gray}{grey}, see section~\ref{sec:spectra} for further details.}
	\label{fig:Neff_plot}
\end{figure}

Figure~\ref{fig:Neff_plot} illustrates in a nutshell the need to consistently combine cosmological and pulsar timing 
information when interpreting the NANOGrav results in terms of a DSPT.
The \textcolor{DESYdunkelblau}{blue} contours show the results of a naive fit of the DSPT parameters --- to be introduced in section~\ref{sec:spectra} ---
to NANOGrav data, {\it without} taking into account physically motivated priors on the rate $\beta/H$
of the phase transition or cosmological constraints.  We discuss the former in more detail in section~\ref{sec:spectra},
and the latter in section~\ref{sec:constraints}. Here, we simply wish to demonstrate that these considerations
(as indicated by grey and red shadings, respectively) will necessarily have a major impact on 
the naively inferred parameter space.
One of our main results from a full statistical treatment, including information from cosmology,
is indeed that an astrophysical explanation of the common red noise signal is much more credible 
than a GWB due to a phase transition from a {\it stable} DS. 
When allowing the DS to {\it decay} at pre-BBN temperatures, on the other hand, we find that
the viable parameter space of DSPTs opens up; in this case, the NANOGrav 
data can be explained without violating BBN constraints, fitting the pulsar timing data as good as SMBHBs.
For earlier work on cosmological constraints on phase transition interpretations of the NANOGrav results, 
see Refs.~\cite{Nakai2020,Bai2021,Deng:2023seh}.

This article is organised as follows.
We start by discussing, in section~\ref{sec:spectra}, how to predict the GW spectra expected from a DS phase transition.
In section~\ref{sec:pta}, we continue with a detailed description of our statistical procedure to analyse PTA data,
remarking also on pitfalls and limitations of simpler or more heuristic methods sometimes adopted in the literature.
We describe the cosmological constraints on DS dynamics  in section~\ref{sec:constraints}, and explain how to construct
global likelihoods simultaneously taking into account both pulsar timing and cosmological information.
We present our results in section~\ref{sec:results}, before concluding in section~\ref{sec:conclusion}.
In three appendices, we provide further technical details about our analysis.

\section{Gravitational wave backgrounds from dark sector phase transitions}
\label{sec:spectra}

The detailed prediction of the GW signal resulting from a first-order phase transition is a highly non-trivial task.
Even though first scaling relations for GWB spectra have been found decades ago~\cite{Kosowsky1992}, 
the form of the GW power spectrum is still subject to active investigations. 
Today, there exist a handful of broadly consistent semi-analytical approximations for predicting GWB spectra emitted by 
first-order phase transitions that are neither too strong nor percolate too slowly~\cite{Caprini2015, Caprini2019}. In deriving these 
spectra, a source of inspiration has often been a detailed investigation of SM extensions that turn the electroweak symmetry 
breaking into a first-order phase transition.

Here we are interested in a DSPT whose characteristic energy scale is unrelated to the electroweak scale. We therefore need to 
adapt the usual approach for calculating GWB spectra from visible sector phase 
transitions~\cite{Huber:2008hg,Hindmarsh:2013xza, Hindmarsh:2015qta, Hindmarsh2017, Cutting:2018tjt, Cutting:2019zws, Jinno:2020eqg, Jinno:2022mie}.
We define a DS as an extension to the SM whose particles are coupled so feebly to the SM species that the two sectors do 
not equilibrate in the early universe. If the couplings {\it within} the DS are sufficiently strong, 
the DS particles instead form a separate thermal bath with temperature $T_\DS$ that is different
from the temperature $T_\SM$ of the SM photon bath~\cite{Breitbach2019}.
In general, the DS temperature ratio $\xi_\DS\equiv T_\DS/T_\SM$ is a time-dependent quantity.
We use $\xi_\DS^\text{perc}$ to denote the ratio of the DS temperature in the old phase to that of  
the SM bath at the time of percolation. Assuming that the energy injection into the DS bath happens instantaneously after 
percolation, this means that $\xi_\DS^\text{perc}$ corresponds to the ratio just \emph{before} the phase 
transition. 
To simplify our analysis, we also assume that the DS reheats instantaneously and that the DS energy density 
after the transition is dominated by at least one relativistic particle species, such that the speed of sound is given by 
$c_\text{s}^\text{DS} = {1}/{\sqrt{3}}$ throughout the transition.\footnote{
The authors of ref.~\cite{Tenkanen2022} found that a 
slightly smaller speed of sound is expected in the broken phase of minimal DS models, which can lead to a sizeable 
suppression of the GW signal for detonations. This would introduce a model dependence which we neglect in our analysis.
}

The amount of vacuum energy released in the transition is encoded in the difference $\Delta \theta$ of the trace of the energy 
momentum tensor between the old and the new phase~\cite{Giese:2020znk, Giese:2020rtr}. 
We can thus introduce
two transition strength parameters
\begin{align}
	\alpha_\tot \equiv  \frac{\Delta \theta}{\rho_\SM^\perc + \rho_\DS^\perc}  &\quad& \text{and} &\quad& \alpha_\DS \equiv \frac{\Delta \theta}{\rho_\DS^\perc} = \alpha_\text{tot} \ba{1 + \frac{g_{\SM,\rho}^\perc}{g_{\DS,\rho}^\perc} \frac{1}{(\xi_\text{DS}^\perc)^4}} \; . \label{eq:alphaDS}
\end{align} 
Here, $\rho_\SM^\perc$ and $\rho_\DS^\perc$ are the energy densities of the SM and DS bath, respectively, at 
the time of percolation (before reheating). They are related through $\xi_\DS^\perc$  
as given above, with $g_{\SM,\rho}^\perc$  ($g_{\DS,\rho}^\perc$) being the number of relativistic SM (DS) energy degrees of 
freedom. While $\alpha_\tot$ determines the amplitude of the GWB spectrum~\cite{Ertas2021}, $\alpha_\DS$ is the 
relevant quantity to use when calculating the efficiency $\kappa(\alpha_\DS, \vw)$ of conversion from vacuum energy to the 
kinetic energy of the bulk plasma motion. To calculate the efficiency factors we use the semi-analytical approximations given in 
the appendix of ref.~\cite{Espinosa:2010hh} for luminal bubble wall velocities $\vw$. 
We note that $\alpha_\DS \gg \alpha_\tot$ for small temperature ratios $\xi_\DS^\perc \ll 1$. 
Hence, the conversion efficiency of vacuum energy to bulk fluid motion is generally large, $\kappa \rightarrow 1$, 
such that 
ultra-relativistic bubble wall velocities $\vw \rightarrow 1$ are obtained.

The gravitational waves emitted during and after a first-order phase transition are sourced by bubble collisions, sound waves, 
and turbulences in the perturbed plasma. We consider the cases of dominant bubble wall collision- and sound wave-induced 
GWB spectra separately. This is a good assumption since bubble wall contributions are dominant for the case of runaway 
bubbles, whereas sound wave contributions dominate in case of bubble walls with terminal velocities. Such a terminal, yet 
ultra-relativistic, velocity exists for example when the symmetry being broken is gauged. In that case, the emission of ultra-soft 
gauge bosons at the bubble walls leads to friction from transition radiation~\cite{Bodeker2017}, i.e.~pressure on the bubble 
wall that grows with the wall's Lorentz factor $\gamma_\text{w}$, unlike the friction from particle kinematics that saturates for 
large velocities~\cite{Bodeker:2009qy}. We do not take into account contributions to the GW spectrum from turbulent motion of 
the DS plasma after the phase 
transition~\cite{Caprini:2007xq, Brandenburg:2017neh, RoperPol:2019wvy, Dahl:2021wyk, Auclair:2022jod} as this source still 
requires a better theoretical understanding.

For relativistic bubble walls the differential energy density in the GWB
today, with respect to the GW frequency $f$, can be parametrized as~\cite{Caprini2019,Huber:2008hg,Ertas2021}
\begin{align}
	h^2 \Omega_\text{GW}(f) &= \mathcal{R}h^2 \times \begin{cases} 
	3 \cdot  \tilde{\Omega}_\sw \, \ba{\frac{\kappa_\sw \, \alpha_\tot}{\alpha_\tot + 1}}^2 \, \ba{8 \pi}^{1/3} \, \ba{\frac{\beta}{H}}^{-1} \, \mathcal{Y}_\text{sh} \times  0.687 \, &s_\sw(f/f_\text{p,\sw})\\
	\tilde{\Omega}_\bw \,  \ba{\frac{\kappa_\phi\, \alpha_\tot}{\alpha_\tot + 1}}^2 \, \ba{\frac{\beta}{H}}^{-2} \, &s_\bw(f/f_\text{p,\bw})
	\end{cases},
	\label{eq:spectra}
\end{align}
 where $\beta$ is the inverse timescale of the transition and $H$ is the Hubble expansion rate.
The first line of the above equation
refers to the case of a GWB dominated by sound wave production and the second line to the case of 
dominant production through bubble wall collisions, for which we adopt spectra normalized to the respective
peak frequencies $f_{p,i}$ as obtained by Refs.~\cite{Hindmarsh2017,Huber:2008hg}, 
\begin{align}
s_\sw(x) = x^3 \ba{\frac{7}{4 + 3 \, x^2}}^{7/2} \qquad \text{and} \qquad s_\bw(x) = \frac{3.8 \, x^{2.8}}{1 + 2.8 \,  x^{3.8}} \;.
\end{align}
At the time of production, the peak frequency is given by $f_{\text{p},\bw}^\text{em}/\beta = 0.23$ for ultra-relativistic
bubble-wall collisions~\cite{Huber:2008hg}, whereas for sound wave-induced spectra it is 
$f_{\text{p},\sw}^\text{em}/\beta = 0.53$~\cite{Hindmarsh2017}. Since the emitted GWs redshift, one has to
instead use 
\begin{align}
	f_{\text{p},i} = \frac{16.5 \, \text{nHz}}{D^{1/3}} \, \ba{\frac{\beta}{H}} \, \ba{\frac{T_\SM^\perc}{100 \, \text{MeV}}} \, \ba{\frac{g_{\tot,\rho}^\perc}{100}}^{1/2} \, \ba{\frac{100}{g_{\tot,s}^\perc}}^{1/3}  \, \ba{\frac{f_{\text{p},i}^\text{em}}{\beta}}\label{eq:peakf}
\end{align}
in eq.~\eqref{eq:spectra}, for the frequencies observed today. Here, we introduced the total entropy 
degrees of freedom $g_{{\rm tot},s}$, in analogy to the energy degrees of freedom $g_{{\rm tot},\rho}$, 
as well as a dilution factor $D\equiv S_{\rm SM}^{0}/S_{\rm tot}^{\rm perc}\simeq1$ 
with $S_{\rm SM}^{0}$ being the SM comoving entropy density today and $S_{\rm tot}^{\rm perc}$ being the total comoving
entropy density just before the transition, respectively
(see below for a discussion)~\cite{Ertas2021,Cirelli2018}.
Further quantities entering in eq.~\eqref{eq:spectra}, to be introduced next, essentially 
fix the overall normalization of the signal.

Starting with the overall prefactor, we use~\cite{Ertas2021}
\begin{align}
\mathcal{R} h^2 &= \frac{h^2 \Omega_\gamma}{D^{4/3}} \, \left(\frac{g_{\text{SM},s}^\text{0}}{g_{\text{tot},s}^\perc}\right)^{4/3} \, \left(\frac{g_{\text{tot},\rho}^\perc}{g_{\gamma,\rho}^\text{0}}\right) = \frac{1.653 \cdot 10^{-5}}{D^{4/3}} \, \ba{\frac{g_{\text{tot},\rho}^\perc}{100}} \ba{\frac{100}{g_{\text{tot},s}^\perc}}^{4/3} \;, \label{eq:redshift}
\end{align}
which is essentially the same as $F_{\text{gw},0}$ in ref.~\cite{Caprini2019}, but taking into account the 
changed redshift history that is induced by the presence of a DS. In the second step we adopted the standard 
$\Lambda$CDM values for today's quantities
$g_{\text{SM},s}^0 = 3.93$, $g_{\gamma,\rho}^0 = 2$, $h^2 \Omega_\gamma = 2.473 \cdot 10^{-5}$, 
and $T_\SM^0 \simeq 2.35 \cdot 10^{-13}\,\text{GeV}$~\cite{Planck2018, Saikawa2018},
noting that the total degrees of freedom at early times can potentially receive large DS 
contributions~\cite{Husdal2016}:
\begin{align}
g_{\text{tot},\rho}^\perc &= g_{\text{SM},\rho}^\perc + g_{\text{DS},\rho}^\perc \, \ba{\xi_\DS^\perc}^4\;, \label{eq:gtotrho}\\
g_{\text{tot},s}^\perc &= g_{\text{SM},s}^\perc + g_{\text{DS},s}^\perc \, \ba{\xi_\DS^\perc}^3 \; .
\end{align}
Next, the factors $\tilde{\Omega}_\sw = 0.012$~\cite{Hindmarsh2017} and 
$\tilde{\Omega}_\bw = 0.11 \, \vw^3 /(0.42 + \vw^2) = 0.077$~\cite{Huber:2008hg} are spectrum normalizations 
obtained in the already previously mentioned simulations; the additional factor of 0.687 ensures that the sound wave spectrum is 
normalized, $\int s_\text{sw}(x) \, \diff \log x  =1/0.687$ (the normalization of the bubble wall spectrum is already absorbed in the 
prefactor $\tilde{\Omega}_\bw$).  
The factor of $\ba{8 \pi}^{1/3}$ in eq.~\eqref{eq:spectra} 
comes from the estimated mean bubble separation~\cite{Caprini2019},
and we explicitly keep the relative factor 3 in the sound wave spectrum pointed out in ref.~\cite{Hindmarsh2017} 
(but erroneously missing in ref.~\cite{Caprini2019}). 
For the case of sound waves, there is an additional suppression of the spectrum by a factor~\cite{Ellis2020, Caprini2019}
\begin{align}
	\mathcal{Y}_\text{sh} = \min \bb{1, \tau_\text{sh} H} \simeq \min \bb{1, \frac{3.38}{\beta/H} \sqrt{\frac{1 + \alpha_\tot}{\kappa_\sw \, \alpha_\tot}}}
\end{align}
if the sound-wave sources last for less than a Hubble time.
Finally, we set $\kappa_\phi = 1$ for bubble wall spectra, while we compute $\kappa_\sw(\alpha_\DS)$ using 
the high-velocity approximation from
ref.~\cite{Espinosa:2010hh}; for large $\alpha_\DS$, this also results in $\kappa_\sw \simeq 1$ 
for the case of non-runaway bubbles.

The duration of the phase transition is parametrised by $\beta/H$, and the two spectra in eq.~\eqref{eq:spectra}
differ in their scaling with this quantity. In particular, the peak amplitude of the bubble wall spectrum
is parametrically suppressed by one power of $\beta/H$, such that sound wave spectra can typically fit a given 
high-amplitude GWB more easily for $\beta/H > 1$.
As the size of this parameter turns out to be very important for the interpretation of our results, let us briefly discuss what is 
physically expected.
The bubble nucleation rate $\Gamma$ is suppressed by the tunnelling action $S_3$. Close to the nucleation time $t_*$ of the phase transition, it is approximately 
\begin{align}
\Gamma &\simeq A \exp \ba{- \frac{S_3(T_\DS)}{T_\DS}} 
\simeq \Gamma_* \, \exp \left[\beta \ba{t-t_*}\right]   \; , \label{eq:nucleationrate}
\end{align}
which using the scaling $\dot{T}_\DS|_{t_*} \simeq - H_* \,  T_\DS^*$ yields
\begin{align}
\beta/H \equiv - \frac{1}{H_*}\left. \frac{\diff}{\diff t}  \frac{S_3(T_\DS)}{T_\DS} \right|_{t_*} 
= T_\DS^* \left. \frac{\diff}{\diff T_\DS}  \frac{S_3(T_\DS)}{T_\DS} \right|_{T_\DS^*} \simeq \left. \ba{S_3/T_\DS}\right|_{t_*} \; .
\end{align}
In the last step we used that quite generally $S_3/T_\DS$ is a polynomial in $T_\DS$, which implies that $\beta/H$ is naturally 
of the same order as $S_3/T_\DS$ at nucleation. The first bubble in a Hubble volume nucleates when 
$\Gamma_* \simeq H_*^4$. Using eq.~\eqref{eq:nucleationrate} one finds that 
$\ba{S_3 / T_\DS} \left.\right|_{t_*} \simeq \log A/H^4_*$. During radiation domination the Hubble parameter is of the order 
$H_*^2 \sim T_\DS^4 / M_\text{pl}^2$ and for dimensional reasons the prefactor in eq.~\eqref{eq:nucleationrate} scales like 
$A \sim T_\DS^4$. Assuming MeV temperatures for the onset of the phase transition, one finds\footnote{
There are contributions of order $\mathcal{O}(4 \log \xi_\DS)$ that are ignored in this estimation, cf.~footnote 1 in 
ref.~\cite{Breitbach2018}. We further ignore the slight difference in the nucleation and percolation temperature for this simple argument.
}
\begin{align}
	\beta/H \simeq \left.\ba{S_3 / T_\DS}\right|_{t_*} \simeq 4 \log M_\text{pl}/T_\DS^* = \mathcal{O}(200) \; . 
\end{align}
Obtaining smaller values for $\beta/H$ in a thermally induced phase transition is possible in near-conformal 
potentials~\cite{Randall:2006py, Konstandin:2011dr, Ellis2020, Kierkla2022} or at the expense of tuning the scalar potential such 
that the system is close to meta-stable (when it will not tunnel during the lifetime of the Universe). 
For much smaller values of $\beta/H\sim\mathcal{O}(1)$, however,
the phase transition might not complete and a phase of eternal inflation might occur. Following the example
of ref.~\cite{Freese2022}, we demand $\beta/H>3$ for successful percolation.
Moreover, the time between the nucleation of the first bubbles to 
percolation is about $10/\beta$ (see, e.g., ref.~\cite{Jinno:2022mie}). 
As simulations neglect the expansion of the Universe during the phase transition, 
which suppresses the GW signal, spectra obtained from 
such simulations are therefore likely overestimated, or at least subject to sizeable uncertainties, for $\beta/H<10$. 

Concerning the exact form of the GW spectra that we adopt here, let us mention that recent results seem to indicate  
sound shell decays leading to an $x^{-3}$ rather than $x^{-4}$ scaling in the UV~\cite{Jinno:2022mie}.
This scaling has little impact on our results since very low phase transition temperatures are disfavoured in our analysis, 
implying that the signal is not fitted by the (far) UV tail. 
Moreover, for wall velocities close to the speed of sound, the 
sound shell thickness (the thickness of the layer that is put into motion for individual bubbles)
becomes imprinted in the fluid motion~\cite{Hindmarsh:2016lnk, Jinno:2020eqg, Jinno:2022mie, Hindmarsh:2019phv}.
This leads to an additional knee in the power spectrum and an intermediate scaling $x^1$.
We also neglect this effect since we focus on wall velocities close to the speed of light.

In all scenarios studied in this work, we set the dilution factor appearing first in eq.~\eqref{eq:peakf} to $D = 1$,  
corresponding to the assumption that there is no significant deviation from the standard cosmological redshift history. 
This is natural in our analysis of stable dark sectors,\footnote{To be precise, the dilution factor $D_\text{SM} = S_\text{SM}^0 / S_\text{SM}^\perc = g_{\text{tot},s}^\perc / g_{\text{SM}, s}^\perc \times D$ is equal to 1 in the stable DS scenario due to comoving entropy conservation \cite{Ertas2021,Cirelli2018}, which corresponds to $D$ being smaller than 1 on the percent level for cold or warm dark sectors phase transitions ($\xi_\DS^\perc < 1$). We neglect this small effect in our calculations.} as the potential dilution gets sourced by an entropy injection, which can only happen for decaying dark sectors. A dilution factor $D > 1$ would correspond to a faster expansion than in radiation domination, e.g.~if the phase transition were followed by an intermediate phase of early matter domination~\cite{Ertas2021}. We checked that this dilution is always negligible also in the case of our decaying dark sector scenario, cf.~appendix~\ref{app:cosmo_decaying_ds}.

We further set $g_{\text{DS},\rho}^\perc = g_{\text{DS},s}^\perc = 1$. This is not a strong assumption, as the DS degrees of freedom always occur as a prefactor of powers of  $\xi_\DS^\perc$ in the calculation. Since the experimentally preferred temperature ratios $\xi_\DS^\perc$ are small, corresponding to cold dark sector phase transitions, the spectrum's dependence on the precise number of DS degrees of freedom is negligible. 
As discussed in section~\ref{sec:constraints}, this choice also does not influence the cosmological constraints
that we impose. 

\begin{figure}[t]
	\centering
	\includegraphics[width=0.85\linewidth]{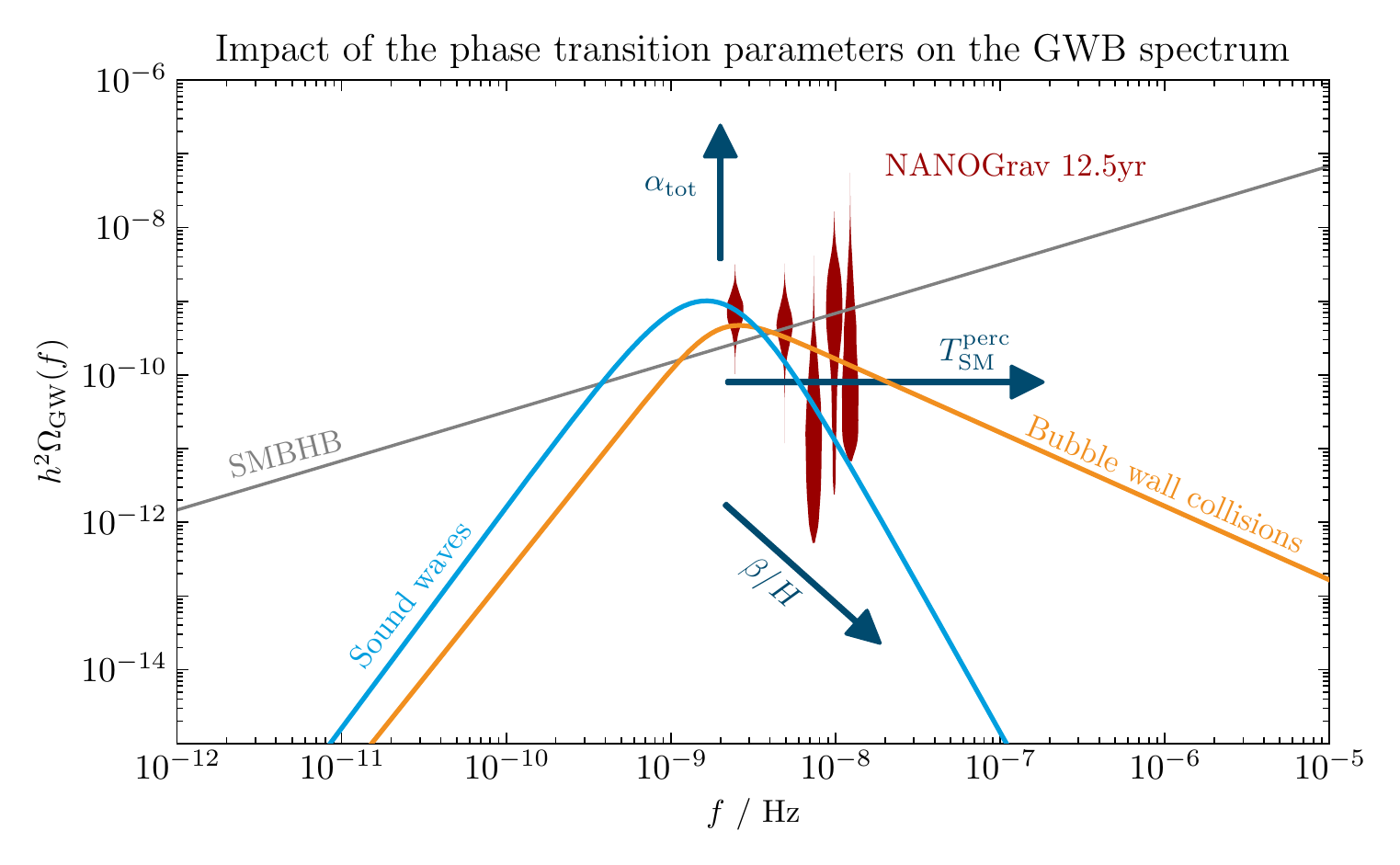}
	\caption{Plot showing the NANOGrav ``violins'' (\textcolor{DESYdunkelrot}{red}) \cite{NANOGrav:2020bcs}, their standard 
	explanation through a power-law spectrum from the inspiral of SMBHBs with $A_\text{SMBHB} = 1.53 \times 10^{-15}$ and 
	$\gamma_\text{SMBHB} = 13/3$ (\textcolor{gray}{grey}) and two phase transition spectra. A characteristic bubble wall 
	collision spectrum is shown in \textcolor{DESYorange}{orange}. A sound wave-induced GWB spectrum is shown 
	in \textcolor{DESYcyan}{blue}. These spectra correspond to the best-fit parameter points found in section~\ref{sec:results}
	(including cosmological constraints and demanding $\beta/H > 1$). The blue arrows indicate 
	how an increase in the phase transition parameters $\alpha_\tot$, $T_\SM^\perc$ or $\beta/H$ would shift the spectra.}
	\label{fig:spectrainfo}
\end{figure}

In summary, we calculate the GWB spectrum based on the following set of DSPT parameters:
\begin{align}
	\bc{\alpha_\tot, \beta/H, T_\text{SM}^\perc, \xi_\DS^\perc} \; .
\end{align}
In figure~\ref{fig:spectrainfo} we illustrate the generic influence of increasing $\alpha_\tot$, $\beta/H$, and 
$T_\SM^\perc$ on the sound wave and bubble wall collision spectra.\footnote{The influence of $\xi_\DS^\perc$ on the GWB spectrum turns out to be negligible in our work since cosmological constraints limit it to a sufficiently small value, cf.~section~\ref{sec:results}.} The spectra shown here
correspond to the best-fit points of the analysis including 
cosmological constraints presented in section~\ref{sec:results} (with a prior of $\beta/H>1$).

\section{PTA data analysis}
\label{sec:pta}

In this section, we briefly go over the methods used to analyse PTA data for common red processes. We start by commenting 
on two often adapted approaches that are used to fit arbitrary GWBs to such signals. After that, 
we discuss in detail why model comparisons based on global fits
require a more rigorous analysis.

In order to fit a given spectrum to the PTA data, one first has to define a deterministic timing model for each pulsar. 
The timing residuals 
are then analysed for ``white'' noise (being uncorrelated in time) and 
time-correlated ``red'' noise. To make the fit to different red noise models computationally feasible, the 
timing model parameters are treated as nuisance parameters and the initial likelihood is marginalized over those.
The remaining likelihood for the low-frequency red 
noise is modelled as a Fourier-series at multiples of the inverse observation period $1/T_\text{obs}$ of the PTA.
Correlations between observed pulsars then help to further distinguish a GWB from the intrinsic noise, as well as  
from other sources of red 
noise (like clock errors or a mismodelled solar system ephemeris). Notably, a GW signal would 
present itself in the data as a common red process before being confirmed through
the characteristic Hellings-Downs cross correlation between the 
pulsars~\cite{Renzini2022,Hellings1983}.

The NANOGrav collaboration has reported strong evidence for a common red noise signal~\cite{NANOGrav:2020bcs}. The 
Bayes factor between this interpretation and the competing no common red noise (nCRN) hypothesis, in which only 
pulsar-intrinsic red noise contributes, exceeds 10.000~\cite{NANOGrav:2020bcs}.  The arguably simplest 
explanation of this observation is a GWB sourced by the inspirals of SMBHBs. In this case the spectrum is expected to follow
a power-law~\cite{Phinney2001},
\begin{align}
	h^2 \Omega_\text{GW}(f) = \frac{2 \, \pi^2}{3 \, H_{100}^2} f^2 \, h_\text{c}^2(f) = \frac{2 \, \pi^2}{3 \, H_{100}^2} A^2_\text{CP} \,  f_\text{yr}^{2} \,  \ba{\frac{f}{f_\text{yr}}}^{5-\gamma_\text{CP}}   \label{eq:PTA_power_law}
\end{align}
with $f_\text{yr} = 1/\text{year}$, $H_{100} = 100 \, \text{km/s/Mpc}$, and the slope $\gamma_\text{CP}=13/3$. Keeping this 
slope fixed, the common process spectral amplitude needs to be $A_\text{CP} = A_\text{SMBHB} \simeq 1.53 \times 10^{-15}$ to 
account for the signal~\cite{NANOGrav:2020bcs}. 

An easy-to-implement possibility to fit arbitrary GWB spectra to the PTA data is to reinterpret the contour plots in the 
$(A_\text{CP}, \gamma_\text{CP})$-plane produced by the experimental collaborations
--- e.g.~figures 1 and 9 in Refs.~\cite{NANOGrav:2020bcs,Antoniadis2022}, respectively  --- 
to also be valid for a GWB whose spectrum is close to a single power-law in a certain frequency interval. 
This method has been used in many 
works~\cite{Freese2022, Benetti2021, Blasi2020, Ellis2020b, Vaskonen2020, Buchmuller2020} that aim to explain the common 
red signal by a GWB from cosmic origins rather than astrophysical sources. 
Since phase transitions result in GWBs with a broken power-law shape,\footnote{
\label{fn:soundshellmodel}In the sound shell model, one expects to see a double broken power-law spectrum with a flat plateau 
shape instead of a distinct peak~\cite{Hindmarsh2020, Giese:2020znk}. This is due to several relevant length scales present in 
the phase transition for slow bubble wall velocities. We stick to the single broken power-law shape in this work, 
cf.~eq.~\eqref{eq:spectra} and figure~\ref{fig:spectrainfo}, motivated by the expected high bubble wall velocities.
} 
this mapping to a specific combination of $A_\text{CP}$ and $\gamma_\text{CP}$ breaks down around the peak of the 
spectrum. While this method is often sufficient for estimating the approximate amplitude of the signal, it is 
not powerful enough for making a proper model comparison between different signal hypotheses.

Another tempting possibility to fit an arbitrary GWB spectrum to the common red noise signal
is to use the results of a free spectral analysis to the PTA data \cite{NANOGrav:2020bcs}. 
In that analysis the assumption of a power-law GWB was dropped in favour of free spectral amplitudes at the 
aforementioned frequencies of integer multiples of $1/T_\text{obs}$. The posteriors of these spectral amplitudes are depicted as 
the infamous ``violins'' reproduced in figure~\ref{fig:spectrainfo}.
Assuming that the total likelihood to fit an arbitrary spectrum to the common red signal factorizes into five 
dominant likelihoods, at the Fourier frequencies $1/T_\text{obs}$, ..., $5/T_\text{obs}$, then 
allows to directly fit spectra that can in principle deviate arbitrarily from a power-law (see, 
e.g., Refs.~\cite{Ratzinger:2020koh,Wang2022, Wang2022a, Ratzinger:2023umd}).
This approach, however, systematically underestimates the uncertainties of the full calculation 
as it by definition neglects correlations between amplitudes at different Fourier 
frequencies~\cite{Ratzinger:2023umd}.
Another significant limitation is that the posteriors for the respective frequency bins are calculated with a finite prior range 
of signal amplitudes. Adding to the fact that the tails of these posteriors are not 
necessarily sampled well enough, this implies that the violins cannot be used in any statistically 
meaningful way for signal amplitudes too far from their most likely values. For example, the 
finite extent of the violins shown in figure~\ref{fig:spectrainfo} would strictly speaking imply that  
the `no-signal' (nCRN) hypothesis is excluded with arbitrary confidence --- while instead it is only disfavoured 
by a Bayes factor of $\sim10^4-10^5$.  

Crucially for us it turns out that cosmological constraints can force a potential DSPT to be so weak that the resulting GWB spectrum only contributes negligibly to the measured common red noise. In that case, the signal is fit by fine-tuning the individual pulsar-intrinsic red noise amplitudes. Since for both the mapping 
to a single power-law and in the free spectral analysis the pulsar-intrinsic red noise components are already marginalized over, it is therefore no longer possible to use one of these quick methods. Instead, in order to treat correlations consistently, a full evaluation of the likelihood is required.

The evaluation of the full PTA likelihood to fit cosmological GWB spectra 
was so far used only in a rather short list of works~\cite{Dandoy2023, NANOGrav:2021flc}, 
due to the large numerical cost of the likelihood evaluations, and --- as far as we are aware of --- never 
in a context that included further constraining likelihoods, like in our case from cosmology.

To interpret the NANOGrav data in terms of a DSPT we first construct a likelihood
$\mathcal{L}_\text{PTA}(\bm{\theta}_\text{PSR}, \bm{\theta}_\DS)$ for fitting the timing residuals to a given set of 
pulsar-intrinsic red noise parameters $\bm{\theta}_\text{PSR}$ and a common red spectrum that depends on the DSPT 
parameters $\bm{\theta}_\DS$. Each pulsar's red noise is fitted by a power-law with an amplitude $A_{\text{red},i}$ and slope 
$\gamma_{\text{red},i}$, in analogy to eq.~\eqref{eq:PTA_power_law}. 
To include constraints from  cosmology on the available dark sector parameter space $\bm{\theta}_\DS$, we further construct a 
likelihood $\mathcal{L}_\text{cosmo}(\bm{\theta}_\DS)$ in section~\ref{sec:constraints}. We multiply this likelihood with the PTA 
likelihood to obtain a global likelihood,
\begin{align}
	\mathcal{L}_\text{glob}(\bm{\theta}_\text{PSR}, \bm{\theta}_\text{DS}) = \mathcal{L}_\text{PTA}(\bm{\theta}_\text{PSR}, \bm{\theta}_\DS) \times \mathcal{L}_\text{cosmo}(\bm{\theta}_\DS) \; . \label{eq:PTA_glob_lik}
\end{align}

In this work we concentrate on the NANOGrav 12.5\,yr data \cite{NANOGrav:2020bcs}, for which 
the full set of arrival time data, the pulsar white noise parameters as well as a 
tutorial on how to use these resources publicly available \cite{NANOGravTut}.\footnote{We plan to include more data sets and the 15\,yr data set of the NANOGrav 
collaboration in a future version of this work.}
In this data set \cite{NANOGrav:2020bcs}, a total of 47 pulsars were taken into account, out of which we consider those 
that were observed for at least three years, as done in the original analysis. From the remaining 45 pulsars, we 
treat the pulsar J1713+0747 as advertised in ref.~\cite{NANOGravTut} due to the probably mis-modelled noise process 
found in the dropout analysis~\cite{NANOGrav:2020bcs}. The parameter space we evaluate therefore consists of 90 
nuisance parameters
$\bm{\theta}_\text{PSR} =\{ A_{\text{red},i}, \gamma_{\text{red},i}\} $ for the pulsar-intrinsic red noise, adding to our four (five)
DSPT model parameters for the case of a (decaying) DSPT. 
To evaluate this high-dimensional global likelihood in a numerically feasible way, we implement the DSPT spectra and
$\mathcal{L}_\text{cosmo}(\bm{\theta}_\DS)$ into the codes \texttt{enterprise} and 
\texttt{enterprise\_extensions}~\cite{enterprise, enterprise2}. To sample 
over the global likelihood we use \texttt{PTMCMC}~\cite{justin_ellis_2017_1037579}. The Markov chain Monte Carlo (MCMC) chains are evaluated using 
\texttt{numpy} and \texttt{scipy}~\cite{harris2020array,2020SciPy-NMeth}, and triangle plots are generated using 
\texttt{matplotlib} and a customized version of \texttt{ChainConsumer}~\cite{Hinton2016}. 
For performing model comparisons, finally,
we calculate Bayes factors using the product space 
method~\cite{10.1093/mnras/stv2217,10.2307/1391010, PhysRevD.91.044048, 10.2307/2346151},
which we briefly review in appendix~\ref{sec:product_space_method}.

\section{Cosmological constraints}
\label{sec:constraints}

The past decades of observational cosmology have provided a large amount of data which allow us to confidently reconstruct 
the cosmological evolution up to MeV-scale temperatures. Most notably these include observations of the CMB, both in terms of 
the spectral shape~\cite{Fixsen:1996nj} and anisotropies~\cite{Planck2018}, and the primordial light element 
abundances produced during BBN~\cite{Workman:2022ynf}. These observations are in very good agreement with the standard 
$\Lambda$CDM model and with each other~\cite{Workman:2022ynf}, implying that any changes to $\Lambda$CDM  
at temperatures below a few $\mathrm{MeV}$ can have observational consequences and need to be 
checked for consistency with available CMB and BBN 
data~\cite{Planck2018,Hufnagel:2018bjp,Forestell:2018txr,Depta2020,Depta:2020mhj,Kawasaki:2020qxm,Yeh2022}.

For a phase transition to produce a strong GW signal a sizeable amount of vacuum energy needs to be released in the 
transition, most of which is subsequently converted into DS energy density as only a small fraction ends up in the GWB. This 
additional energy density could change the well-tested cosmic expansion history, 
possibly even long after the transition itself. To understand whether NANOGrav may observe the remnants of 
such a phase transition we therefore need to include a cosmological likelihood $\mathcal{L}_\text{cosmo}$ when analysing the 
PTA data. 
Specifically, we include information from the primordial light element abundances and CMB anisotropies into our analysis as 
described below.\footnote{
Note that constraints from $\mu$-distortions of the CMB photon spectrum~\cite{Ramberg2022} and curvature 
perturbations~\cite{Liu2022} are less relevant as they quickly lose sensitivity for transition temperatures above the MeV-scale. 
PBH formation due to first-order phase transitions~\cite{Lewicki2023,Gouttenoire2023,Baker2021} could offer novel probes, but 
turns out to be irrelevant for the phase transition strengths of interest in our work.
}

\subsection{Stable dark sector}

If the entire DS energy density after the phase transition is contained in
light degrees of freedom, this contributes an 
additional radiation energy density that can be described by a (potentially large) additional contribution $\DNeff$ to the effective 
number of neutrinos $N_\text{eff} = N_\text{eff}^\text{SM} + \DNeff$, where $N_\text{eff}^\text{SM} = 3.044$~\cite{Bennett2020} 
is the SM prediction for $\Lambda$CDM cosmology.\footnote{
The assumption of a radiation-dominated DS is conservative in the sense that constraints only become tighter 
if the energy density instead starts to redshift as matter at some time after the phase transition. We also note that the 
contribution to $\DNeff$ from the GWs themselves, for a GWB with peak amplitude 
$h^2 \Omega_\text{GW}^\text{peak} \lesssim 10^{-9}$ as required to explain the 
NANOGrav signal, is typically less than $\DNeff \simeq 10^{-3}$, cf.~ref.~\cite{Maggiore:2018sht}, and therefore irrelevant in 
the discussion of cosmological constraints.
} 
The effective number of neutrinos affects the predictions of BBN as well as the CMB power spectrum and is constrained by 
observations to $N_\text{eff} = 2.941 \pm 0.143$~\cite{Yeh2022}. This bound can be modelled by a Gaussian likelihood $\mathcal{L}_{\text{BBN} + \text{CMB}} (N_\text{eff})$, i.e.
\begin{align}
\mathcal{L}_\text{cosmo} (\bm{\theta}_\DS) = \mathcal{L}_{\text{BBN} + \text{CMB}} (N_\text{eff} = N_\text{eff}^\text{SM} + \DNeff(\bm{\theta}_\DS))\;,
\end{align}
enabling us to implement the cosmological constraints for the case of a stable DS. 
The number of degrees of freedom in the DS does not have any relevant impact on cosmological constraints as the available latent 
heat would simply be distributed among the different species, leaving the total injected energy density unchanged.\footnote{
Taking into account the energy density \emph{before} the phase transition, a change in the number of degrees of freedom can be 
absorbed in the temperature ratio $\xi_\mathrm{DS}$ which we treat as a free parameter in our scans anyway.} We therefore set 
$g_{\DS,\rho}^\perc=1$ in our analysis.
For more details we refer to appendix~\ref{app:cosmo_stable_ds}.

\subsection{Decaying dark sector}
If there are additional small inter-sector couplings --- which happens very naturally due to possible `portal' couplings such as 
Higgs mixing for scalars or kinetic mixing for dark photons ---  the energy density injected into the DS will subsequently be 
transferred to the SM heat bath via decays of DS particles. In this case cosmological constraints in general depend on the 
lifetime, mass, and coupling structure of the decaying particles. 
As a simple concrete example and for minimality, we consider the DS after the phase transition to consist only of one bosonic 
degree of freedom $\phi$ decaying into photons or electrons with a lifetime $\tau_\phi$ that we sample over. This is a natural 
setup as a light scalar degree of freedom with mass below the phase transition temperature is generally expected to exist and 
e.g.~mixing with the Higgs is generally allowed. Given that the NANOGrav data prefer an MeV-scale phase transition temperature, also the mass of $\phi$ is 
expected to be of this order, $m_\phi \simeq \mathrm{MeV}$. 
In the example of Higgs mixing very short lifetimes $\tau_\phi$ correspond to a 
sizeable Higgs mixing angle $\theta$ which is constrained by a number of laboratory experiments to be $\theta \lesssim 10^{-4}$ for small masses~\cite{Winkler:2018qyg,Ferber:2023iso}. 
Translating this constraint for MeV-scale masses into the lifetime results in $\tau_\phi \gtrsim 10^{-2} \; \mathrm{s}$ whereas cosmological constraints roughly require $\tau_\phi \lesssim 10^{-1} \; \mathrm{s}$, so that some allowed region remains in this scenario.\footnote{Note that the case of a Higgs-mixed scalar is particularly constrained because of the Yukawa-suppressed couplings to electrons, implying a rather long lifetime. If the relevant state would be e.g.~a dark photon, the allowed range of lifetimes would be significantly larger.}
As already indicated, the resulting cosmological constraints largely depend on the lifetime 
$\tau_\phi$ of the particle with only a very mild dependence on the mass as long as the energy density in $\phi$ is not very 
suppressed.\footnote{Smaller masses
generally lead to strong constraints for arbitrarily short lifetimes $\tau_\phi$ due to thermalisation by inverse decays.} To simplify 
our analysis we therefore fix the mass of the decaying particle to $m_\phi = 5 \; \mathrm{MeV}$,
noting that the results will be very similar for other choices of the mass in the MeV-range. 
To implement the cosmological constraints, we use a likelihood incorporating BBN and the CMB power spectrum constructed 
from the results of~\cite{Depta2020,Bai2021} as detailed in appendix~\ref{app:cosmo_decaying_ds}.

\section{Results}
\label{sec:results}

We now discuss the results of our work, based on global fits of the model setups described in the previous sections. We start by
considering phase transitions in a stable dark sector (section~\ref{sec:results_stable}), and then turn to  
a decaying dark sector that thermalises with the visible sector some time after the phase transition (section~\ref{sec:results_decay}).
We finally compare different GW explanations of the signal in terms of their respective Bayes factors, 
in section~\ref{sec:results_comp}, and discuss how future data might strengthen the DSPT interpretation.

\subsection{Stable dark sector phase transitions}
\label{sec:results_stable}

Let us first focus on GWs that are mainly emitted as a consequence of the bulk motion of the DS plasma 
after the phase transition, i.e.~a GWB dominantly produced through sound waves. 
The full set of model parameters  to describe such a scenario for a stable DSPT, as introduced in section~\ref{sec:spectra}, 
is given by
\begin{align}
	\{\alpha_\tot, \beta/H, T_\SM^\perc, \xi_\DS^\perc\}\;.
\end{align}
We sample over these input parameters with flat log priors, as well as over 90 nuisance parameters $\bm{\theta}_\text{PSR}$
for the pulsar-intrinsic red noise, based on the combined PTA and cosmological likelihood given in eq.~\eqref{eq:PTA_glob_lik}.
For a full overview over parameters and prior ranges, see also table~\ref{tab:priors} in appendix~\ref{sec:priors}.

\begin{figure}[t]
	\centering
	\includegraphics[width = 0.85\textwidth]{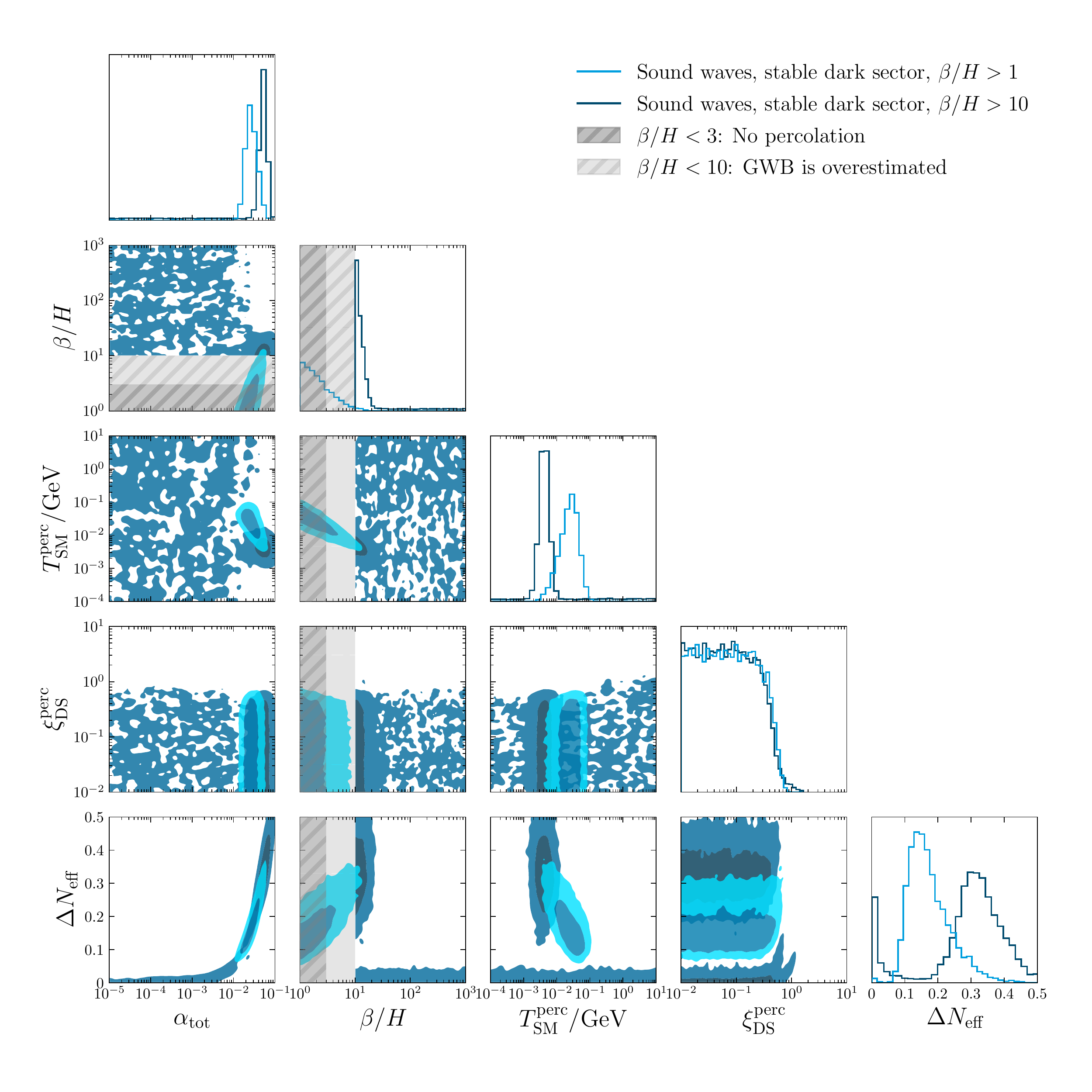}
	\caption{Comparison of the triangle plots for two MCMC chains with different priors on $\beta/H$, assuming 
	a stable DS and a GWB generated through sound waves. Blue shaded regions mark $1\,\sigma$ and $2\,\sigma$ contours, respectively. Regions with $\beta/H<3$ and 
	$\beta/H<10$, in which the phase transition cannot complete and the GW signal is overestimated respectively, are  indicated with \textcolor{gray}{grey} shadings.} 
	\label{fig:tringle-comparison}
\end{figure}

We show the resulting  corner plot of posterior distributions for the four model parameters in figure~\ref{fig:tringle-comparison}, 
to which we  add the derived parameter $\DNeff$. 
Allowing for inverse time scales down to $\beta/H > 1$ (\textcolor{DESYcyan}{light blue}) formally 
results in a good global fit of the pulsar timing residuals, as indicated by the compact ellipsoid-like posterior regions 
where the NANOGrav signal is explained by the GWB.  Such small values of the transition rate would however suppress the GW spectrum w.r.t.~the commonly adopted parametrization, cf.~the discussion in section~\ref{sec:spectra}, and, for $\beta/H \lesssim 3$, likely not even lead to successful percolation. We therefore
also show, in the same figure, the results of a fit with a more restrictive prior of $\beta/H > 10$
(\textcolor{DESYdunkelblau}{dark blue}). 
In this case, the best-fit region moves to a somewhat larger value of $\alpha_{\rm tot}$,
but it also 
becomes apparent that there is no longer a single preferred region in the model parameter space. Instead, 
the combined data now shows a similar preference for a very weak DSPT-induced GW signal with correspondingly weak 
cosmological constraints, where the NANOGrav signal is not explained by the GWB but absorbed in the pulsar-intrinsic noise 
parameters. This indicates that also the best-fit region no longer corresponds to an equally 
plausible interpretation of the combined data set.

The reason is that cosmology adds a constraint on $\DNeff$ which effectively translates into a constraint on 
the phase transition strength $\alpha_{\rm tot}$. In terms of fitting the  NANOGrav signal, the required lower value of $\alpha_{\rm tot}$ can partially 
be compensated by a lower value of $\beta/H$ (see also figure~\ref{fig:spectrainfo}).
And indeed, comparing the posterior distributions for $\beta/H$ in figure~\ref{fig:tringle-comparison}, we can see that 
$\beta/H$ always sticks closely to the lower prior 
boundary --- which is qualitatively different from the analysis without cosmological constraints, 
cf.~figure~\ref{fig:Neff_plot}, where inverse timescales of $\beta/H = \mathcal{O}(10-100)$ are favoured.
Increasing the lower prior bound on $\beta/H$ therefore  directly induces a shift towards larger  
$\alpha_{\rm tot}$. At the same time, a larger inverse timescale $\beta/H$ also means smaller 
bubbles at the time of their collision, and hence a larger peak frequency in the spectrum
(see again figure~\ref{fig:spectrainfo}).
This is compensated for by a  lower percolation temperature $T_\SM^\perc$, which by itself would lead to smaller peak frequencies.
Finally, we can identify  in figure~\ref{fig:tringle-comparison} an upper bound on the initial  temperature ratio, which again
 is a direct consequence of the constraint on $\DNeff$. For $\xi_\DS^\perc \gtrsim 0.7$, in particular, eq.~\eqref{eq:Neff} 
 would imply a violation of this constraint already for a single dark sector species that is not non-relativistic before the 
 transition~\cite{Breitbach2018}.

Overall, these effects push the posterior for $\DNeff$ towards higher values. Since this is strongly punished by the cosmological
part of the likelihood, however, that also explains the already mentioned appearance of a second preferred parameter region,
characterized by a weak DSPT that corresponds to $\DNeff \simeq 0$ (at the price of an unobservably small GW signal). 
We confirmed that this region of parameter space is indeed explained by fine-tuning the pulsar-intrinsic red noise 
parameters, rather than by a GWB, by directly comparing the marginalized posteriors of the nuisance 
parameters $\bm{\theta}_\text{PSR}$ between the two chains depicted in \textcolor{DESYcyan}{light blue} 
and \textcolor{DESYdunkelblau}{dark blue}. Let us stress that this parameter range would
have been impossible to reliably infer with simpler statistical methods, i.e.~by re-fitting
 a power-law common red spectrum described by $\ba{A_\text{CP}, \gamma_\text{CP}}$ or by using the 
 free-spectrum ``violins'' (see the discussion in section~\ref{sec:pta}).
 
\begin{figure}[t]
	\centering
	\includegraphics[width = 0.8\textwidth]{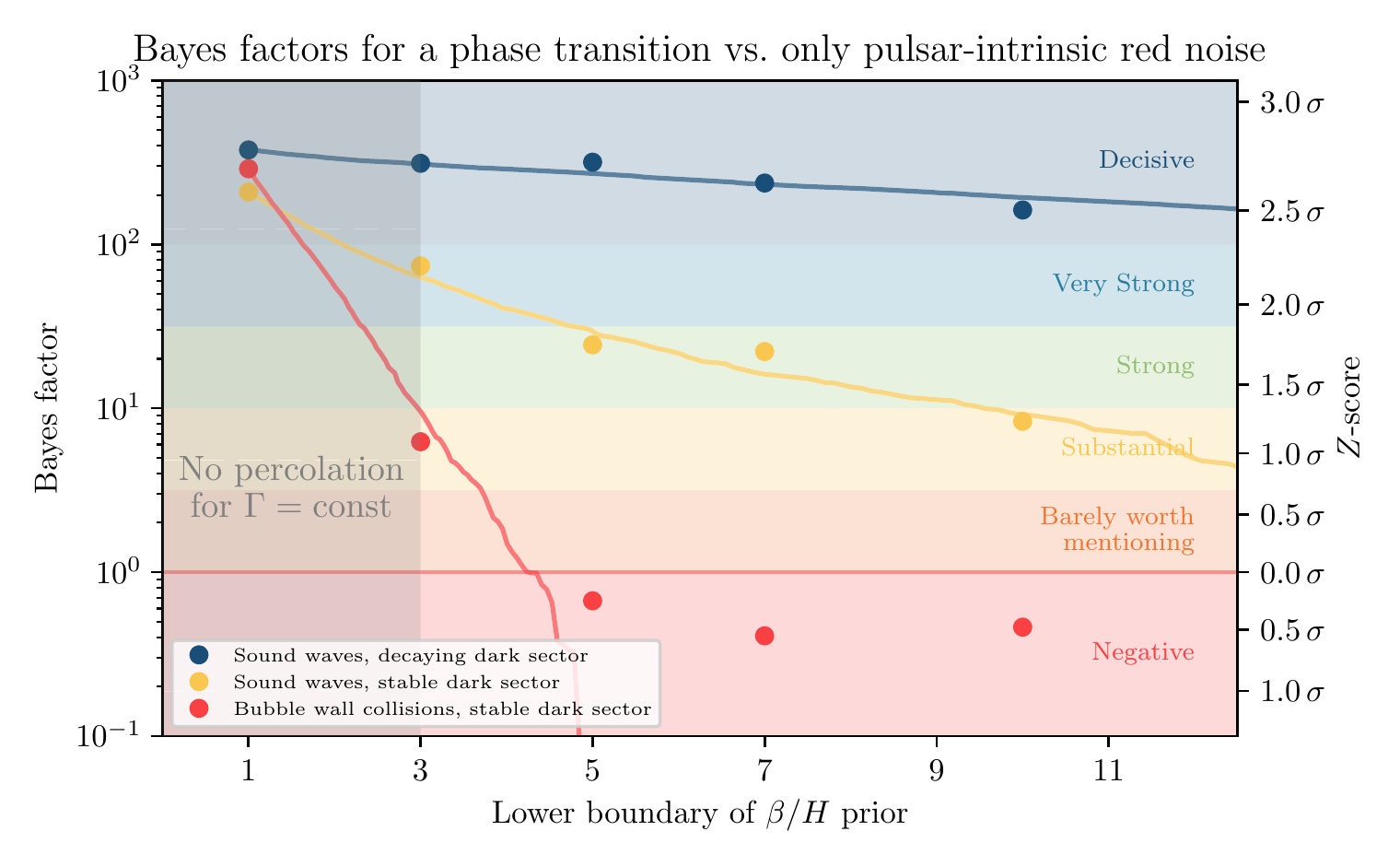}
	\caption{Bayes factor estimates of various DSPT models with respect to the no common red noise-hypothesis. 
	Filled coloured dots correspond to actually performed full model comparisons, while lines in the corresponding colours are 
	derived from an {\it a posteriori} reduction of the prior range of chains with minimal $\beta/H$ of 1 (for details, see 
	appendix~\ref{sec:priorchoice}). We also indicate how to translate the Bayes factor to  Jeffrey's scale (coloured shadings)
	as well as $Z$-scores from frequentist statistics (right $y$-axis, cf.~appendix~\ref{sec:z-scores}). 
	In the grey shaded area ($\beta/H < 3$), a constant nucleation rate is not sufficient to drive
	percolation.}
	\label{fig:BFs}
\end{figure}

From the above discussion, one would expect the Bayes factor between the DSPT and the no nCRN 
hypothesis to decrease when increasing the lower prior bound on $\beta/H$, as higher and higher values of $\DNeff$  
are needed to explain the combined data. We confirm this expectation explicitly in figure~\ref{fig:BFs}.
Here, each of the coloured dots corresponds to a separate MCMC chain that was employed to determine the Bayes factor 
between the two models by using the product space method explained in appendix~\ref{sec:product_space_method}. 
The corresponding lines serve as a cross-check for the prior dependence of the Bayes factors, 
see appendix~\ref{sec:priorchoice} for further details.
Yellow dots and lines refer to the case of a stable DSPT where the GWB production is dominated by sound waves;
this corresponds to the same model as shown in figure~\ref{fig:tringle-comparison}.

For comparison, we also show the
case of a GWB that is mostly due to bubble wall collisions (red). Just from the point of view of the resulting spectrum,
cf.~figure~\ref{fig:spectrainfo}, one might expect that this could be a viable alternative. Compared to sound waves,
 however, bubble wall spectra receive an additional parametric suppression of $h^2\Omega_\text{GW}^\text{peak}$ 
 by a factor of $\ba{\beta/H}^{-1}$. This induces the need for a larger $\alpha_{\rm tot}$ and hence an even stronger
 constraint on $\DNeff$, making the GWB hypothesis worse than the nCRN assumption for $\beta/H\gtrsim5$.
 To further illustrate these considerations, we refer to figure~\ref{fig:spec-comparison} in appendix~\ref{sec:posteriorGWBdist}, showing
 the posterior distribution of the bubble wall spectra for different prior choices on  $\beta/H$.
Note also that neither figure~\ref{fig:tringle-comparison} nor figure~\ref{fig:BFs} include the expected suppression in the GWB 
spectra for $\beta/H\lesssim10$, which would further decrease our Bayes factor estimates.

Overall we therefore come to the conclusion that a stable DSPT can hardly compete with the alternative SMBHB explanation of 
the NANOGrav timing residuals, once one takes into account cosmological constraints from BBN and CMB. 
For $\beta/H>10$, in particular, the Bayes factor between a DSPT explanation and the nCRN hypothesis is 
only $\mathcal{O}(10)$ even in the favourable case of sound wave-induced GWBs ---
much smaller than the factor of $\sim10^{4.5}$ that is claimed for a GWB from SMBHBs~\cite{NANOGrav:2020bcs}. 

\subsection{Decaying dark sector phase transitions}
\label{sec:results_decay}
We next consider a DS that couples sufficiently strongly to ordinary matter such that it can decay 
after the phase transition. A decay long before BBN, in particular, is not subject to the cosmological constraints
that played such a decisive role for the case of a stable DSPT.
On the other hand, a phase transition happening too early would produce a GWB at too high frequencies
to be compatible with the NANOGrav signal. It therefore is a non-trivial question whether any relevant parameter
space remains where PTA and cosmological data are indeed compatible.
For concreteness, we assume the decay of a dark Higgs boson as detailed in section~\ref{sec:constraints}
that decays with a lifetime $\tau_\phi$, such that our model parameters read 
\begin{align}
\{\alpha_\tot, \beta/H, T_\SM^\perc, \xi_\DS^\perc, \tau_\phi\} \; .
\end{align}
For the dark Higgs lifetime we adopt a log prior ranging from $10^{-6}\, \text{s}$ to $10^2 \, \text{s}$; the remaining
parameters we treat as in the previous section (with $\beta/H>1$).

\begin{figure}[t]
	\centering
	\includegraphics[width=0.85\linewidth]{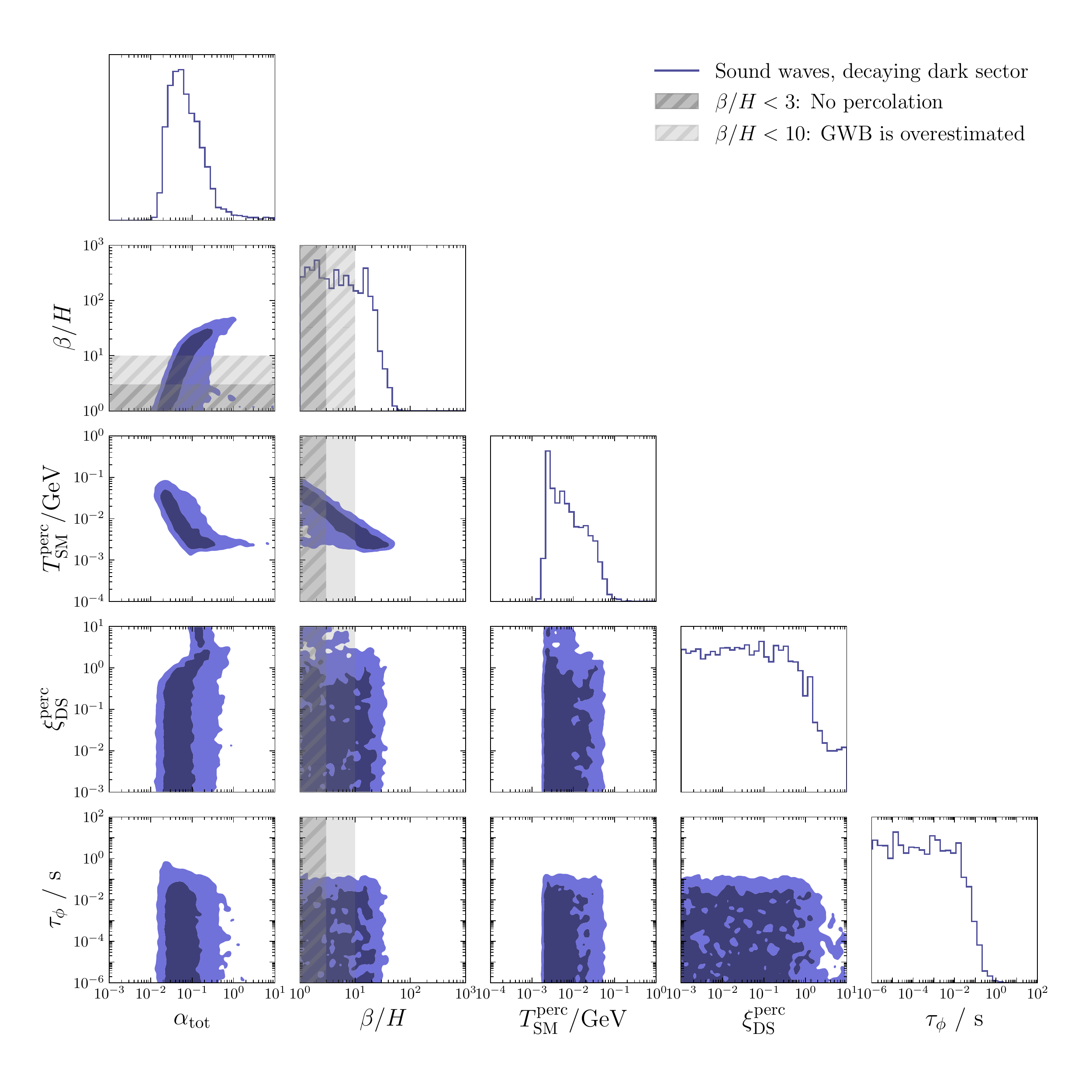}
	\caption{Posterior distributions of model parameters in the decaying dark sector scenario.
	}
	\label{fig:dsdecay_logprior}
\end{figure}

We show the resulting triangle plot for this model in figure~\ref{fig:dsdecay_logprior}. 
In a nutshell, we find that the common red spectrum observed by NANOGrav can be explained as long as 
$\tau_\phi \lesssim 0.1 \, \text{s}$ and $T_\text{SM}^\perc \gtrsim 2 \, \text{MeV}$.
Larger lifetimes, corresponding to temperatures smaller than 2\,MeV, 
are strongly constrained as the decays occur after the onset of BBN or neutrino decoupling. Such decays change the time-temperature relation of the SM heat bath 
and alter the ratio of the neutrino and photon temperatures, leading to a {\it negative} contribution to $\DNeff$.
If the percolation temperature $T_\SM^\perc$ drops to values close to the temperature of neutrino decoupling strong constraints arise independent of the lifetime $\tau_\phi$.
Note that we implemented the results from ref.~\cite{Bai2021} for simplicity as a sharp cut-off enforcing 
$T_\SM^\perc>2 \, \text{MeV}$ in our likelihood, cf.~appendix~\ref{app:cosmo_decaying_ds}. Adopting a more accurate 
likelihood would result in a smoother transition of the posterior in the range $T_\SM^\perc \approx 1 - 2 \, \text{MeV}$, 
the main effect being a slight increase of the maximal possible value of $\beta / H$.

Compared to our analysis of stable DSPTs, the posterior for the inverse timescale $\beta/H$ is relatively flat up to 
$\beta/H\sim30$, implying a very limited prior dependence. The underlying reason for this is
 the possibility of dumping the liberated energy density into the SM photon bath before the neutrino decoupling at 
 around $2 \, \text{MeV}$, thereby evading cosmological constraints and hence opening up for large phase transition 
 strengths $\alpha_\tot\gtrsim0.1$ to fit the common red signal even for $\beta/H \gtrsim 10$. 
 This however only works up to the point when $\beta/H$ becomes so large that its effect on the peak frequency can no 
 longer be compensated for by a correspondingly lower percolation temperature, cf.~figure~\ref{fig:spectrainfo}.

In figure~\ref{fig:BFs} we also indicate the Bayes factor for the decaying DSPT scenario (blue). As anticipated, the 
prior dependence on $\beta/H$ is much less severe than in the scenarios discussed previously.
In particular, this shows that a GWB from a decaying DSPT is a viable explanation of the common red noise signal
even for $\beta/H > 10$. Quantitatively, the model evidence is a factor of $\sim200$ larger than that of the 
nCRN hypothesis, corresponding to $2.5\, \sigma$ or a ``decisive'' evidence on Jeffrey's scale.

\subsection{Comparison with other GW sources}
\label{sec:results_comp}

Let us next address in more detail the question of how a DSPT interpretation of the signal compares to alternative 
GWB hypotheses, in particular the leading astrophysical explanation of an SMBHB-induced GWB, and how future data
will help to further distinguish these two.

We start by pointing out that the DSPT spectra actually fit the common red spectrum in the NANOGrav data slightly {\it better} 
than the SMBHB spectra. Naively, this is already expected from figure~\ref{fig:spectrainfo}, and we demonstrate this
in more detail in appendix~\ref{sec:priorsDSPT}. Nevertheless the maximal Bayes factor (with respect to nCRN) that 
we find is only about $10^{2.5}$, significantly smaller than the $\sim10^{4.5}$ typically quoted for SMBHBs. 
We checked explicitly that the reason is not connected to the goodness of fit, but entirely due to prior volume effects:
Bayesian statistics dutifully renders the simple SMBHB explanation of the data more credible than the 
apparently more complicated DSPT model.
It is however important to keep in mind that the amplitude $A_\text{SMBHB}$
of the astrophysical signal is not necessarily 
a single fundamental  parameter, as assumed in deriving a Bayes factor of $10^{4.5}$, but effectively derived from 
more fundamental parameters in a more complicated astrophysical model,
such as astrophysical merger timescales or the mass-, redshift-, and spatial distributions of the SMBHBs. 
A full Bayesian analysis should thus also consider constraints on these fundamental parameters, as e.g.~done in 
ref.~\cite{Casey-Clyde2021} without fitting the NANOGrav data.
This would decrease the formal evidence for the SMBHBs interpretation because {\it (i)} the prior volume increases 
due to the intrinsic parameters that the amplitude depends on, and {\it (ii)} 
astrophysical models predict amplitudes that are smaller than
those inferred from the NANOGrav data \cite{Casey-Clyde2021,Kelley2016, Kelley2017, Antoniadis2022}.
Analogously, of course, the DSPT parameters $\alpha_\tot$, $\beta/H$, $T_\SM^\perc$, $\xi_\DS^\perc$, and $\tau_\phi$ 
are in reality derived quantities from a given SM extension whose independent parameters are masses and couplings. Evaluating specific models where these underlying parameters are known would be interesting and clearly deserves further study, but is not the aim of our more model-independent analysis.

Concerning near-future perspectives, the first thing to be confirmed is that NANOGrav indeed sees a GWB
by verifying that the correlations follow a Hellings-Downs curve~\cite{Hellings1983}. This is is expected to happen
relatively soon, within the next weeks to years~\cite{Pol2020}, and thus possibly already for the pending release 
of the 15\,yr data set.
Even with a confirmed GW signal one may initially expect the {\it relative} odds between SMBHBs and an 
alternative DSPT explanation to remain about the same --- as long as the preferred spectral shape of the GWB
remains similarly vaguely determined as of today. This is simply because, as discussed above, the prior volume is currently
the main driver to distinguish between these two hypotheses.
In other words, the curves in figure~\ref{fig:BFs} will likely all shift towards higher Bayes factors in favour of DSPTs,  because
the relative pull by $\DNeff$ towards lower values of $\alpha_\tot$ will reduce, but the Bayes factors 
for the SMBHB interpretation will improve by roughly the same factor.

A first lever-arm to really distinguish the DSPT from the SMBHB scenario will be a determination of
the spectral index $\alpha=(3-\gamma)/2$.
This quantity could be determined by up to $40\,\%$ relative accuracy
once Hellings-Downs is confirmed~\cite{Pol2020}. Measuring a value of $\gamma$ that is at odds with the
value of $\gamma_\text{SMBHB} = 13/3$ expected for SMBHBs~\cite{Phinney2001} would then be a strong indicator
for a DSPT instead of an astrophysical interpretation of the GWB. 
Current data, in fact, already slightly prefer $\gamma>13/3$, cf.~figure~\ref{fig:spectrainfo}.
This could easily be accommodated by the GWB expected from a DSPT, both close to the 
peak amplitude and in the high-frequency tail of the spectrum.
An even clearer signature would be a broken power-law, but to test for such a feature will require even
higher signal-to-noise ratios and hence take correspondingly  longer time to probe observationally. 
A valuable source of information to further disambiguate between SMBHBs and GWBs of primordial origin  
may lie in the anisotropies of the GWB~\cite{Schulze2023,Bartolo2022,Taylor2020}, in particular via the 
polarization of the GWB~\cite{Kato:2015bye,Conneely:2018wis,Hotinli:2019tpc,Belgacem:2020nda,Sato-Polito:2021efu,ValbusaDallArmi:2023ydl,Ellis:2023owy} and possibly the detection of singular GW sources as already searched 
for~\cite{NANOGrav:2023bts,IPTA:2023ero}.

Let us finally stress 
that not only the evidence for a nHz GWB will change in the future; also the competing cosmological 
constraints on phase transitions are subject to improvements. Upcoming experiments like the Simons observatory and CMB-S4 
measurements in combination with surveys of large-scale structure will be able to push the limits on $\DNeff$ by about an order 
of magnitude~\cite{Abazajian2019, Simons2018, Dvorkin2022}, contributing to the tension on the stable dark sector explanation 
we investigated above (and reducing the parameter space for a decaying DSPT). Measurements of 
CMB spectral distortions with PIXIE~\cite{Chluba2019} will give additional and complementary constraints~\cite{Ramberg2022}, 
which would be relevant even in the case of a decaying dark sector that avoids constraints on $\DNeff$. 
In fact, a confirmation of spectral
distortions in the CMB could in principle even provide {\it supporting} evidence for such a scenario.
It will therefore remain crucial to include all relevant cosmological observables as well as 
new PTA data to eventually decide on the most probable origin of a putative GWB signal. 

\section{Conclusions}
\label{sec:conclusion}

We investigated the appealing possibility that the common red spectrum observed by the NANOGrav collaboration in their 
12.5\,yr data set~\cite{NANOGrav:2020bcs} is due to a dark sector phase transition (DSPT) just before the onset of Big Bang 
Nucleosynthesis (BBN). For the first time, we performed a global analysis on pulsar timing array (PTA) data from a (potential) 
gravitational wave background (GWB) including constraints from BBN and the cosmic microwave background (CMB) 
anisotropies.

We found that a dark sector (DS) undergoing a phase transition can in principle explain the measured signal with a goodness of 
fit that is comparable to, or even better than, that of the standard astrophysical explanation in terms of a stochastic GWB from 
supermassive black hole binaries (SMBHBs). However, if one accounts for
\begin{enumerate}
	\item the changes in the early element abundances that the energy density released during 
	the phase transition would induce,
	\item the impact on the CMB anisotropies trough a contribution to $\DNeff$, and
	\item possible issues for transitions with $\beta/H< 10$, connected to percolation and an overestimation of the produced GWB,
\end{enumerate}
the possibility of a {\it stable} DSPT no longer gives a good fit to all available data.
Figure~\ref{fig:Neff_plot} provides an intuitive illustration of this tension, 
by directly confronting the above constraints with the results of 
a naive DSPT analysis of the NANOGrav data that ignores them.
Fully including all relevant constraints in a global fit, the available parameter space is 
indeed significantly reduced, cf.~figure~\ref{fig:tringle-comparison}.

On the other hand, there is no intrinsic reason why a DS should be stable on cosmological  timescales.
In particular, tiny interactions with the visible sector (e.g.~through small portal couplings~\cite{Batell:2009di})
could well lead to a decay before neutrino decoupling at $T_\SM^\perc \gtrsim 2 \, \text{MeV}$. 
We find that such a {\it decaying} DSPT scenario remains a compelling alternative to the more conventional 
SMBHB hypothesis for lifetimes $\tau_\phi \lesssim 0.1 \, \text{s}$, cf.~figure~\ref{fig:dsdecay_logprior}.
We arrived at this conclusion by further taking into account constraints on electromagnetic energy injection from decaying dark 
scalars~\cite{Depta2020} and on the reheating of the photon bath after a phase transition~\cite{Bai2021}.
Compared to the no-GWB hypothesis, we find a Bayes factor that indicates a decisive evidence for the DSPT interpretation
even for a prior of $\beta/H>10$ on the transition rate. The currently maximal value of this quantity that is compatible
with the data, $\beta/H\lesssim50$,  
still indicates the need for a relatively slow transition; further model-dependent research will be needed to 
investigate how this can be implemented in a given SM extension.

We also studied the effect of prior choices on the absolute scale of Bayes factors, finding that prior volume effects are
highly relevant when comparing SMBHB and DSPT explanations of the NANOGrav data. 
The SMBHB interpretation, in particular, seemingly only requires one parameter to fit the signal, namely the amplitude 
$A_\text{SMBHB} \simeq 1.53 \times 10^{-15}$. We however argue that
$A_\text{SMBHB}$ 
should rather be treated as a derived quantity that depends on several intrinsic, independently measured astrophysical 
quantities~\cite{Casey-Clyde2021}. This would reduce the difference between the Bayes factors above $10^4$ for the SMBHB 
explanation~\cite{NANOGrav:2020bcs} and the Bayes factors of $\mathcal{O}(10^2)$ that we find for the decaying DSPT
interpretation.

We remain excited about the pending 15\,yr data release of the NANOGrav collaboration, which might already confirm the 
Hellings-Downs curve needed to state the detection of a GWB~\cite{Hellings1983, Pol2020}. While we do not expect a definite 
answer concerning the origin of the signal within the coming months, we are confident that additional PTA data as well 
as complementary constraints from cosmology~\cite{Ramberg2022} will help to disambiguate between different models. 
The earliest evidence we can hope for in favour of a cosmological origin of the signal will be any deviation from the spectral slope 
of $\gamma_\text{SMBHB} = 13/3$ expected for SMBHBs. With further data it will be ever more crucial to  
fully include complementary constraints such as from BBN and CMB 
when assessing different signal models.
In this work we have made a first step in this direction,
thereby contributing to moving the realm of testable cosmology to pre-BBN times.

\acknowledgments
We would like to thank Xiao Xue, Andrea Mitridate, and Felix Kahlh\"ofer for helpful discussions on related works. This work is funded by the Deutsche Forschungsgemeinschaft (DFG) through Germany's Excellence Strategy – EXC 2121 ``Quantum Universe'' --- 390833306. CT is grateful for the hospitality of the Swedish Collegium for Advanced Study during the very final stage of this work.

\newpage
\appendix	
\addtocontents{toc}{\protect\setcounter{tocdepth}{1}}

\section{Details on cosmological likelihood}
\label{app:details_cosmo}
	
In this appendix we provide details on the computation of the relevant quantities for the cosmological likelihoods as well as the mapping to previously published results.
	
\subsection{Stable dark sector}
\label{app:cosmo_stable_ds}

For the model of a stable DS after the phase transition we assume that the energy density in the DS after the transition is contained purely in radiation. Note that constraints would only be more stringent if the DS energy density would start to redshift like matter, increasing the ratio to the SM energy density. Furthermore, we focus on transitions before the onset of BBN and neutrino decoupling. At the observationally relevant temperatures we can then model the DS energy density by an additional contribution $\DNeff$ to the effective number of neutrinos $N_\text{eff}$. This assumption is validated a posteriori by the results of our MCMC chains.

At the time of percolation $t_\perc$ the DS energy density in radiation $\rho_\DS^\perc$ can be quantified through the effective number of relativistic degrees of freedom $g_{\DS, \rho}^\perc$ and the DS temperature $T_\DS^\perc = \xi_\DS^\perc \, T_\SM^\perc$,
\begin{align}
	\rho_\DS^\perc = \frac{\pi^2}{30} \, g_{\DS, \rho}^\perc \, (\xi_\DS^\perc \, T_\SM^\perc)^4 \; .
\end{align}
Assuming an instantaneous reheating of the DS after the phase transition, one finds
\begin{align}
	\rho_\text{DS}^\text{reh} = \rho_\text{DS}^\perc + \Delta V \; , \label{eq:cosmo_instant_reheating}
\end{align} 
where the index reh ($\perc$) indicates a point in time immediately after (before) the DS reheating. The strength of the phase transition is characterized by $\alpha_\tot \equiv \Delta \theta / \ba{\rho_\text{DS}^\perc +\rho_\text{SM}^\perc}$, where $\Delta \theta \simeq \Delta V$ for sufficiently strong transitions. We thus find
\begin{align}
\rho_\text{DS}^\text{reh} =  \left(1 + \alpha_\tot\right)\rho_\text{DS}^\perc + \alpha_\tot \, \rho_\text{SM}^\perc \quad 
\Rightarrow \quad \frac{\rho_\text{DS}^\text{reh}}{\rho_\text{SM}^\perc} = \alpha_\tot +  \left(1 + \alpha_\tot\right)\frac{\rho_\text{DS}^\perc}{\rho_\text{SM}^\perc} \label{eq:rho_ds_reh_gen}
\end{align}
In terms of the temperature ratio $\xi_\text{DS}^\perc$ before the phase transition we obtain
\begin{align}
\frac{\rho_\text{DS}^\text{reh}}{\rho_\text{SM}^\reh} &= \alpha_\tot +  \left(1 + \alpha_\tot\right)\frac{g_{\text{DS},\rho}^\perc}{g_{\text{SM},\rho}^\perc} \ba{\xi_\text{DS}^\perc}^4 \label{eq:energyratioreh} \; .
\end{align}
Energy density in radiation additionally to the SM can be quantified by its contribution to the late time effective number of relativistic neutrino species
\begin{align}
	N_\text{eff} &= \frac{8}{7} \left(\frac{11}{4}\right)^{4/3} \frac{\rho_\text{DS} + \rho_\nu}{\rho_\gamma}\;.
\end{align}
By subtracting the SM prediction in a $\Lambda$CDM cosmology $N_\mathrm{eff}^\SM = 3.044$~\cite{Bennett2020} we can define the contribution of the DS to the effective number of relativistic neutrino species
\begin{align}
	\Delta N_\text{eff} = N_\text{eff} - N_\text{eff}^{\SM} = \frac{8}{7} \left(\frac{11}{4}\right)^{4/3} \frac{\rho_\text{DS}}{\rho_\gamma}  \; .
\end{align}
Plugging in eq.~\eqref{eq:energyratioreh} gives
\begin{align}
	\Delta N_\text{eff} &\simeq  \frac{4}{7} \left(\frac{11}{4}\right)^{4/3} \left(\frac{3.93}{g_{\text{SM},s}^{\perc}}\right)^{4/3} g_{\text{SM},\rho}^{\perc} \left[\alpha_\tot +  \left(1 + \alpha_\tot\right)\frac{g_{\DS,\rho}}{g_{\text{SM},\rho}^{\perc}} (\xi_\text{DS}^\perc)^4\right] \; , \label{eq:Neff}
\end{align}
where we inserted the SM degrees of freedom today and assumed the DS degrees of freedom $g_{\DS,\rho} = g_{\DS,\rho}^\perc$ to remain constant. This reproduces eq.~(1) from ref.~\cite{Bai2021}.

Observationally, $N_\text{eff}$ affects the predictions of BBN as well as for the CMB power spectra. Combining data sets from the Planck satellite with observations of the primordial abundances of deuterium and helium-4 and marginalizing over the baryon-to-photon ratio $\eta = n_\text{b} / n_\gamma$, ref.~\cite{Yeh2022} finds $N_\text{eff} = 2.941 \pm 0.143$. We approximate this by a Gaussian likelihood.\footnote{By using the likelihood directly in our calculations we do not need to discuss the difference between one-sided ($\DNeff \geq 0$) or two-sided ($\DNeff \in \mathbb{R}$) limits. The former simply result from a one-sided integration of the likelihood starting from $N_\text{eff}^\text{SM}$, cf.~ref.~\cite{Yeh2022}.}
\begin{align}
	\mathcal{L}_\text{cosmo}(\alpha_\tot, \xi_\DS^\perc, T_\SM^\perc) = \frac{1}{\mathcal{N} \sqrt{2 \pi \sigma_{N_\text{eff}}^2}} \exp \bb{ - \frac{\ba{\Delta N_\text{eff} (\alpha_\tot, \xi_\DS^\perc, T_\SM^\perc) + N_\text{eff}^\text{SM} - \mu_{N_\text{eff}}}^2}{2 \, \sigma_{N_\text{eff}}^2} } \; , \label{eq:stableLik}
\end{align}
where $\mu_{N_\text{eff}} = 2.941$, $\sigma_{N_\text{eff}} = 0.143$, $N_\text{eff}^\text{SM} = 3.044$, and the normalization needed for the correct determination of Bayes factors is given by
\begin{align}
\mathcal{N} = \frac{1}{\sqrt{2 \pi \sigma_{N_\text{eff}}^2}} \exp \bb{ - \frac{\ba{N_\text{eff}^\text{SM} - \mu_{N_\text{eff}}}^2}{2 \, \sigma_{N_\text{eff}}^2} }\;.
\end{align}
As $\DNeff$ is simply a function of the phase transition parameters, we can now evaluate the cosmological likelihood for a DS phase transition with given parameters. By multiplying this likelihood to the one for the timing residuals we can simultaneously fit the NANOGrav data and check for consistency with cosmological observables.

\subsection{Decaying dark sector}
\label{app:cosmo_decaying_ds}

The constraints from BBN and the CMB on a decaying particle with $\mathrm{MeV}$-scale mass have been calculated in ref.~\cite{Depta2020} for a general setup. The cosmological evolution was started at an SM temperature\footnote{The subscript $\mathrm{cd}$ refers to chemical decoupling, having a setup of thermal production of dark matter by freeze-out (chemical decoupling) in a DS in mind. Note that the temperature ratio in ref.~\cite{Depta2020} was denoted by $\zeta$ instead of $\xi$.} $T_\SM^\mathrm{cd} = 10 \, \mathrm{GeV}$ and corresponding time $t_\mathrm{cd}$ with a Bose-Einstein distribution function of $\phi$ of temperature $T_\DS^\mathrm{cd} = \xi_\DS^\mathrm{cd} \, T_\SM^\mathrm{cd}$ and zero chemical potential, i.e.
	\begin{align}
	f_\phi^\mathrm{cd} (p) = \left[\exp \left(\frac{\sqrt{p^2 + m_\phi^2}}{\xi_\DS^\mathrm{cd} \, T_\SM^\mathrm{cd}}\right) -1\right]^{-1} \; .
	\end{align}
The corresponding number density is
\begin{align}
n_\phi^\mathrm{cd} = \int \frac{\mathrm{d}^3 p}{(2 \pi)^3} f_\phi^\mathrm{cd} (p)\;. \label{eq:cosmo_decds_n_cd}
\end{align}
To use the results and constraints from ref.~\cite{Depta2020} we therefore need to map to this scenario.

Sufficiently strong intra-sector couplings, i.e.~such that the interaction rate is much larger than the Hubble rate, lead to a self-thermalisation of the DS quickly after the phase transition. If the processes are further $\phi$-number violating, we can expect the chemical potential of $\phi$ to vanish. The phase-space distribution function after the DS has reheated at $t_\mathrm{reh}$ can hence be described by a Bose-Einstein distribution function
	\begin{align}
	f_\phi^\reh (p)  = \left[\exp \left(\frac{\sqrt{p^2 + m_\phi^2}}{\xi_\DS^\reh \,  T_\SM^\reh}\right) -1\right]^{-1} \label{eq:fphi_reh}
	\end{align}
with corresponding number and energy densities
\begin{align}
n_\phi^\reh &= \int \frac{\mathrm{d}^3 p}{(2 \pi)^3} f_\phi^\reh (p)\;,\label{eq:cosmo_decds_n_reh} \\
\rho_\phi^\reh &= \int \frac{\mathrm{d}^3 p}{(2 \pi)^3} \sqrt{p^2 + m_\phi^2} \, f_\phi^\reh (p)\;. \label{eq:cosmo_decds_rho_reh}
\end{align}
To compute the energy density $\rho_\phi^\DS$ we again assume an instantaneous reheating with $\rho_\DS^\perc = (\pi^2/30) \, g_{\DS,\rho}^\perc \, (\xi_\DS^\perc \, T_\SM^\perc)^4$ and compute (cf.~eq.~\eqref{eq:rho_ds_reh_gen})
\begin{align}
\rho_\phi^\reh &= \rho_\DS^\reh \nonumber \\
&= (1 + \alpha_\tot ) \, \rho_\text{DS}^\perc + \alpha_\tot \, \rho_\text{SM}^\perc \nonumber \\
&= \frac{\pi^2}{30} \, g_{\SM,\rho}^{\perc} \, \ba{T_\SM^\perc}^4  \left[\alpha_\tot +  \left(1 + \alpha_\tot\right)\frac{g_\DS^\perc}{g_{\text{SM},\rho}^{\perc}} (\xi_\text{DS}^\perc)^4\right] \; . \label{eq:dec_ds_rho_phi_reh}
\end{align}
For fixed mass and lifetime, cosmological constraints are mostly driven by the comoving number density of a particle. We hence compute the comoving number density $n_\phi^\reh$, which can be obtained from $\rho_\phi^\reh$ by numerically solving eq.~\eqref{eq:cosmo_decds_rho_reh} for $T_\DS^\reh = \xi_\DS^\reh \, T_\SM^\perc$ and then computing $n_\phi^\reh$ from eq.~\eqref{eq:cosmo_decds_n_reh}. This last step of mapping $\rho_\phi^\reh$ to $n_\phi^\reh$ can be accelerated considerably by utilizing tabled data for zero chemical potential and fixed mass $m_\phi$. By redshifting number densities using comoving conservation of SM entropy density $s_\SM$,
\begin{align}
n_\phi^\mathrm{cd} = n_\phi^\reh \frac{s_\SM^\mathrm{cd}}{s_\SM^\reh} = n_\phi^\reh \frac{g_{\SM,s}^\mathrm{cd}}{g_{\SM,s}^\reh} \left( \frac{T_\SM^\mathrm{cd}}{T_\SM^\reh} \right)^3\;, \label{eq:dec_ds_mapping}
\end{align}
and then inverting $n_\phi^\cd$ numerically to obtain the required temperature ratio $\xi_\DS^\cd$, cf.~eq.~\eqref{eq:cosmo_decds_n_cd}, we can map the case of a DS phase transition to the case computed in~ref.~\cite{Depta2020}. We comment below on the validity of this mapping.

To construct a likelihood, we use the calculated primordial light element abundances, their theoretical errors from nuclear rate uncertainties, and $N_\text{eff}$ underlying figure~6 (left) of ref.~\cite{Depta2020}. These are compared to the recommended values of the observed primordial light element abundances of deuterium $\mathrm{D / {}^1H}^\obs = (2.547 \pm 0.025) \times 10^{-5}$ and the mass fraction of helium-4 $\mathcal{Y}_p^\obs = (2.45 \pm 0.03) \times 10^{-1}$~\cite{Workman:2022ynf} as well as $N_\text{eff}^\obs = 2.99 \pm 0.17$~\cite{Planck2018}. As the BBN calculations use the baryon-to-photon ratio $\eta$ (or equivalently the baryon density in the Universe) from ref.~\cite{Planck2018} as an input, but $N_\text{eff}$ is affected by the DS decays injecting energy in the photon heat bath, we use the best-fit value of $\eta$ for given $N_\text{eff}$ from figure~26 (Planck TT, TE, EE+lowE+lensing+BAO) of ref.~\cite{Planck2018}. Note that due to the strong dependence of the prediction of $\mathrm{D / {}^1H}$ on $\eta$ we need to propagate the uncertainty in the determination of $\eta$, effectively marginalizing over $\eta$, such that the total observational error becomes $\eta_{\mathrm{D / {}^1H}}^\obs = 0.035 \times 10^{-5}$~\cite{Depta2020}. After using the $N_\text{eff}$-dependent best-fit value there still remains the aforementioned constraint on $N_\text{eff}$. The cosmological likelihood is given by a product of Gaussian likelihoods with mean at the observed values and total errors obtained by summing the observational and the theoretical error (negligible for $N_\mathrm{eff}$) in quadrature.

In the calculation above we started from eq.~\eqref{eq:dec_ds_rho_phi_reh}, crucially assuming that the DS and the SM only thermalise after reheating, i.e.~that (inverse) decays of $\phi$ only become relevant after reheating. This holds as long as $\tau_\phi > t_\reh$, where $t_\reh$ can be obtained in radiation domination via the Hubble rate as $H(T_\SM^\reh) \simeq 1/(2 t_\reh)$. For shorter lifetimes $\tau_\phi < t_\reh$ the DS and the SM thermalise during or shortly after reheating. Therefore, the temperature ratio after reheating is generally (close to) one, $\xi_\DS^\reh = 1$, but $T_\SM^\perc$ is no longer equal to $T_\SM^\reh$ as reheating goes into the DS as well as the SM due to the decays of $\phi$. We find $T_\SM^\reh = T_\DS^\reh$ by solving
\begin{align}
\rho_\SM^\reh + \rho_\phi^\reh &= (1 + \alpha_\tot) (\rho_\SM^\perc + \rho_\DS^\perc) \label{eq:cosmo_ds_dec_implicit_T_SM_reh}
\end{align}
for $T_\SM^\reh$ with $\xi_\DS^\reh = 1$ in eq.~\eqref{eq:fphi_reh}. With $T_\SM^\reh$, $\rho_\phi^\reh$, and $\xi_\DS^\reh = 1$ we can find $n_\phi^\reh$ as before and map to the results of ref.~\cite{Depta2020} using eq.~\eqref{eq:cosmo_decds_n_cd}. This still assumes that there is no change in comoving number density of $\phi$ or comoving SM entropy density in the calculation of ref.~\cite{Depta2020} between the temperature $T_\SM^\mathrm{cd} = 10 \, \mathrm{GeV}$ and the numerical value of $T_\SM^\reh$, i.e.~decays and inverse decays can be neglected up to the corresponding numerical value of the time $t_\reh$. Strictly speaking, this assumption is not valid any more due to thermalisation around a time $t \sim \tau_\phi < t_\reh$. However, successful thermalisation erases all knowledge of initial conditions, implying that we only have an inaccurate mapping if thermalisation itself has observable consequences, i.e.~it occurs during BBN. Hence, also the reheating process would need to occur during BBN, and we, in fact, have to include the effect of the reheating process (also reheating the SM) on BBN. These were studied in ref.~\cite{Bai2021} (under the assumption $\xi_\DS^\perc = 0$), finding that $\alpha_* \gtrsim 0.07$ is constrained if $T_\SM^\perc \lesssim 2 \, \mathrm{MeV}$ for reheating into photons. To compare this to our case we note that in ref.~\cite{Bai2021} the transition strength is defined by $\alpha_* = \Delta \theta / \rho_\SM^\perc$ such that $\rho_\SM^\reh = (1 + \alpha_*) \, \rho_\SM^\perc$ and we need to map
\begin{align}
\alpha_* = \alpha_\tot + (1 + \alpha_\tot) \frac{g_\DS^\perc}{g_\SM^\perc} (\xi_\DS^\perc)^4 - \frac{\rho_\phi^\reh}{\rho_\SM^\perc}\;. \label{eq:alphaast}
\end{align}
The complete cosmological likelihood is given by
\begin{align}
&\mathcal{L}_\text{cosmo}(\tau_\phi, \alpha_\tot, \xi_\DS^\perc, T_\SM^\perc) \nonumber \\
&= \frac{1}{\mathcal{N}} \times \mathcal{L}_{\mathcal{Y}_p} \times \mathcal{L}_{\mathrm{D / {}^1H}} \times \mathcal{L}_{N_\mathrm{eff}} \times \begin{cases}
1 \; &\text{for} \; \tau_\phi > t_\reh \\
\theta(\max[0.07 - \alpha_*, T_\SM^\perc - 2 \; \mathrm{MeV}]) \; &\text{for} \; \tau_\phi < t_\reh
\end{cases} 
 \;,
\end{align}
where the normalization is $\mathcal{N} = \mathcal{L}_{\mathcal{Y}_p} \times \mathcal{L}_{\mathrm{D / {}^1H}} \times \mathcal{L}_{N_\mathrm{eff}}$ for standard $\Lambda$CDM cosmology with only the SM contributing to the energy density during BBN.

The likelihood given above accurately describes the relevant cosmological constraints on a decaying scalar in a quite model-independent way. We however needed to make some assumptions along the way, either to assure the numerical feasibility of our calculations or to keep the number of parameters describing the decaying dark sector scenario low to allow for a phenomenological interpretation of our results. Note that these assumptions generally do not affect our conclusions that the NANOGrav signal can be explained well by a decaying DSPT as long as the energy from the DS is injected into the SM before the onset of BBN and neutrino decoupling (i.e.~$\tau_\phi \lesssim 0.1 \, \mathrm{s}$ and $T_\SM^\perc \gtrsim 2 \, \mathrm{MeV}$). Still, we discuss the effect of the assumptions below for completeness.

A first simplification is the choice of a particular mass $m_\phi = 5 \, \mathrm{MeV}$ due to readily available data from ref.~\cite{Depta2020}. Generally, the dependence on the mass is expected to be very mild. Only when the mass is small enough, $m_\phi \lesssim 2 \, \mathrm{MeV}$, for the decaying particle to thermalise with the SM heat bath as a relativistic particle for $\tau_\phi \lesssim 0.1 \, \mathrm{s}$ and act as an additional relativistic degree of freedom, arbitrarily small lifetimes can be constrained~\cite{Depta2020}.

Next, we use results for decays into photons. The results of ref.~\cite{Depta2020} show that decays into electron-positron pairs give very similar constraints for $m_\phi>2\, m_e$. For $m_\phi < 2 \, m_e$ decays into electron-positron pairs are kinematically forbidden such that a corresponding coupling will, in fact, mostly lead to decays into two photons and the constraints \textit{on the total lifetime} are just as in the case for decays into photons (whereas constraints on couplings would become suppressed by loops and additional SM couplings). Other interesting portals of the two sectors that lead to a thermalisation between them are imaginable, e.g.~due to the kinetic coupling of a dark photon field with the SM photon~\cite{Heeba2019}. These couplings are however beyond the scope of our work due to the additional model-dependent constraints.

When considering a particular DS model, the assumption of instantaneous reheating might not be justified, e.g.~if thermalisation of the DS is not sufficiently fast due to smaller intra-sector couplings or if there is additional dilution of the relevant energy densities between $t_\perc$ and $t_\reh$ in eq.~\eqref{eq:cosmo_instant_reheating} due to a longer reheating process.

By only redshifting the number density of the decaying particle via eq.~\eqref{eq:dec_ds_mapping} we neglected a possible change in comoving number density due to decays and inverse decays in the calculation of ref.~\cite{Depta2020} between the starting temperature $T_\SM^\mathrm{cd} = 10 \, \mathrm{GeV}$ and the numerical value of $T_\SM^\reh$. While our treatment using the results of ref.~\cite{Bai2021} redeems this shortcoming adequately, a full model-dependent study should calculate the whole cosmological evolution of the DS including all relevant energy transfers between the DS and the SM. In this way, also the dilution factor $D$ entering the GWB spectrum would be calculated~\cite{Ertas2021}, which we simply set to 1. This assumption is valid as long as the decaying particle does not become non-relativistic for an extended period of time before its decay. We checked a posteriori how large the dilution factor would be for the regions favoured by NANOGrav data, cf.~figure~\ref{fig:dsdecay_logprior}, using the \texttt{dilution} sub-package of \texttt{TransitionListener} from ref.~\cite{Ertas2021}. Within the $1 \, \sigma$ contour, the dilution factor deviates by at most $2 \, \%$ from 1. The maximal dilution factor within the $2 \, \sigma$ contour is $D \approx 2$. Hence, our choice of setting $D = 1$ is not only conservative but also well justified in the relevant parameter space.

Finally, our assumption of a negligible chemical potential might be violated in some setups. As long as number-changing processes are efficient, the chemical potential is driven to zero. However, if the decaying scalar is not completely relativistic after reheating and the scalar self-couplings are low, number-changing processes can cease to be relevant and a chemical potential can develop~\cite{Bringmann:2020mgx,Hufnagel2022}. Such a scenario would likely require a full numerical solution of the respective Boltzmann equations and is therefore beyond the scope of this work.

\section{Posterior distribution of GWB spectra}
\label{sec:posteriorGWBdist}

\begin{figure}[t]
	\centering
	\includegraphics[width = \textwidth]{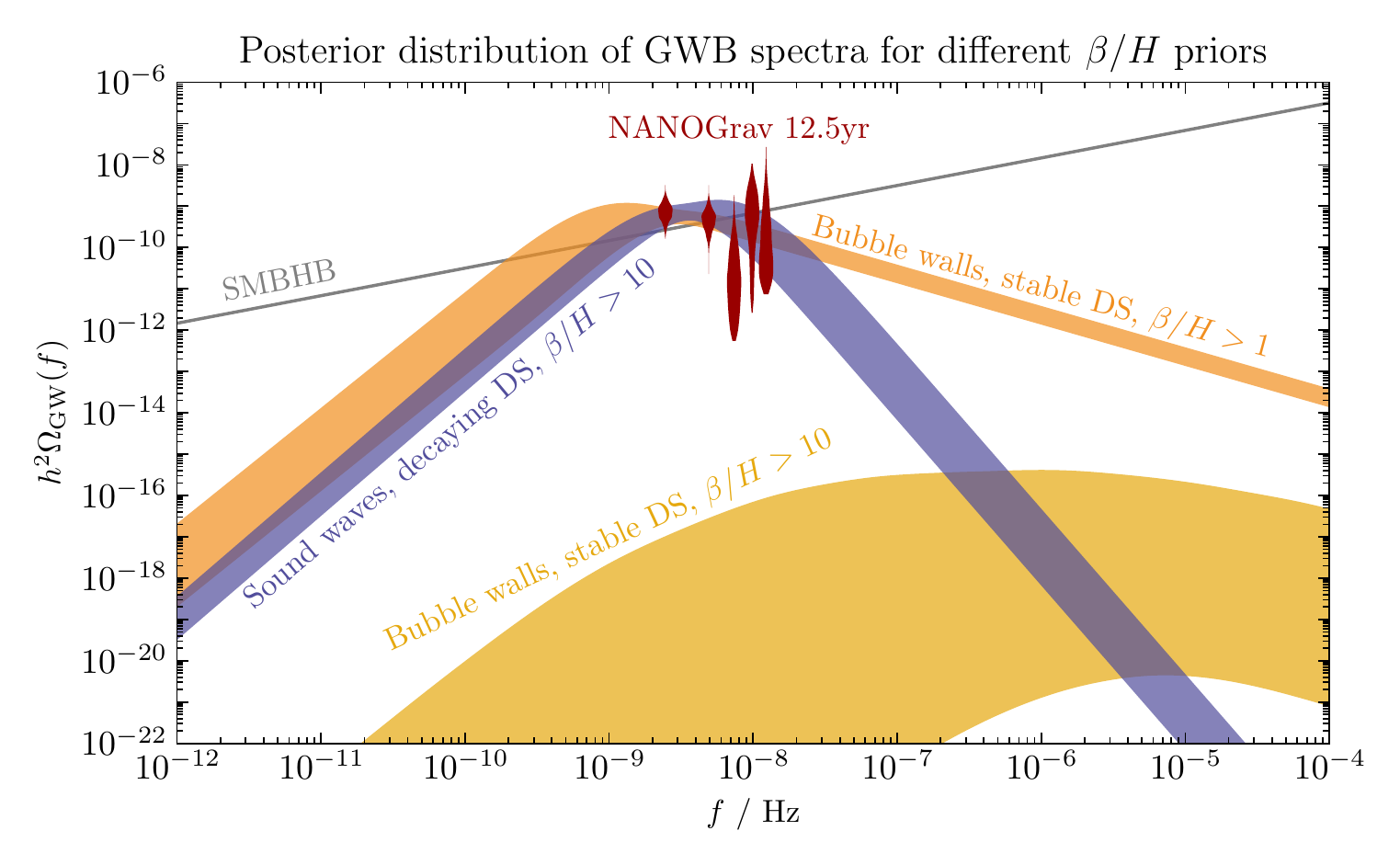}
	\caption{Posterior distributions of GWB spectra for parameter scans assuming bubble collision spectra from a stable DS phase transition with $\beta/H$ above $1$ (\textcolor{DESYorange}{orange}) and above $10$ (\textcolor{DESYgelb}{yellow}). Due to cosmological constraints, the spectra with $\beta/H > 10$ clearly cannot fit the common red spectrum detected by NANOGrav~\cite{NANOGrav:2020bcs} (cf.\ \textcolor{DESYdunkelrot}{red} violins). If the dark sector can decay (\textcolor{DESYlila}{violet} spectra), slightly faster transitions with $\beta/H > 10$ can still explain the observed signal. The GWB from SMBHB with $A_\text{SMBHB} = 1.53 \times 10^{-15}$~\cite{NANOGrav:2020bcs} is depicted as a \textcolor{gray}{grey} line for comparison.}
	\label{fig:spec-comparison}
\end{figure}

To further demonstrate the possible tension between cosmological constraints and the interpretation of the NANOGrav data in terms of a DS phase transition, we illustrate the posterior distribution of GWB spectra for different parameter scans in figure~\ref{fig:spec-comparison}.\footnote{The GWB posterior distributions are illustrated using the mean and standard deviation of 1000 spectra randomly drawn from their respective posterior distributions. The shaded areas hence correspond to the $1 \sigma$ preferred spectral amplitudes.} The \textcolor{orange}{orange} curves show the distribution of bubble wall collision spectra from a dark sector phase transition with $\beta/H>1$. For bubble wall collision spectra with $\beta/H>10$ (\textcolor{DESYgelb}{yellow} curves), the tension in the data becomes apparent through the low signal amplitudes. The spectra cannot explain the common red spectrum, as depicted by the \textcolor{DESYdunkelrot}{red} ``violins'', due to the strong constraints from $\DNeff$. When going to a decaying DS (\textcolor{DESYlila}{violet}), a phase transition with $\beta/H>10$ can clearly still be consistent with the measured signal.

This visualization also clearly shows that a mere fit using the violins is not sufficient for our analysis including cosmological constraints due to the unknown likelihood for small GWB spectra, where the signal is absorbed in finely tuned pulsar-intrinsic red noise parameters, cf.~section~\ref{sec:pta}.

\section{Details on the calculation of Bayes factors}

	\subsection{The product space method}
		\label{sec:product_space_method}
		The calculation of Bayes factors in this work relies on the product space method, which we briefly review for completeness. To compare two models, a hyper-model $\mathcal{H}$ is introduced, whose parameter space contains the Cartesian product $\bm{\theta}_\mathcal{H}$ of the two sub-models' parameters $\bm{\theta}_0$ and $\bm{\theta}_1$ as well as an additional model index $n$ that can run from $-0.5$ to $+1.5$. The key idea of this method is that the model index $n$ can be treated as an additional, continuous parameter that is sampled over in an MCMC chain. At any given step of the evaluation of the hyper-model, the underlying algorithm casts the model index to either 0 or 1, corresponding to one of the two sub-models. Given a current model index $n$, the hyper-model parameter space is partitioned in an active part, which is used to evaluate the respective likelihood of the sub-model, and an inactive part. The posterior odds ratio $\mathcal{P}_{01}$ is then be relative amount of chain entries in model 1 compared to model 0, from which the Bayes factor $\mathcal{B}_{01}$ between the two models can be deduced.
		
		To show how this procedure can be used to calculate a Bayes factor consider the posterior probability for the model index $n$
		\begin{align}
			p(n| \text{data}, \mathcal{H}) = \int \text{d}\bm{\theta}_\mathcal{H} \, p(\bm{\theta}_\mathcal{H}, n  | \text{data}, \mathcal{H}) = \frac{1}{\mathcal{Z}_\mathcal{H}} \int \text{d}\bm{\theta}_\mathcal{H} \, p(\text{data} | \bm{\theta}_\mathcal{H}, n, \mathcal{H}) \, p(\bm{\theta}_\mathcal{H}, n | \mathcal{H}) \; . \label{eq:BF_prob_n}
		\end{align}
		The first equality used that the posterior for $n$ can be obtained by marginalizing the posterior for $n$ and $\bm{\theta}_\mathcal{H}$ over the hyper-model parameters. Bayes' theorem leads to the second equality, where $\mathcal{Z}_\mathcal{H}$ is the hyper-model evidence, the first integrand is the likelihood and the second one is the prior for the model parameters and the model index itself. The hyper-model evidence is unknown and difficult to obtain, but of no further importance, as we are interested in the posterior odds ratio $\mathcal{P}_{01}$ between the two models. For a fixed $n$ we can factorize the prior $p(\bm{\theta}_\mathcal{H}, n | \mathcal{H})$ in eq.~\eqref{eq:BF_prob_n} as
		\begin{align}
			p(\bm{\theta}_\mathcal{H}, n | \mathcal{H}) = p(\bm{\theta}_n| \mathcal{H}_n) \, p(\bm{\theta}_{\overline{n}} | \mathcal{H}_{\overline{n}}) \, p(n | \mathcal{H}) \; . \label{eq:BF_prior}
		\end{align}
		This factorization uses the aforementioned distinction between active (first factor) and inactive (second factor) parameters, which are not correlated with each other as the sub-models are distinct. The last factor is a subjective prior for each of the sub-models. The two last factors do not depend on the active parameters. If we insert these expressions into the definition for the posterior odds ratio for the model index, we find 
		\begin{align}
			\mathcal{P}_{01} &= \frac{p(n=1| \text{data}, \mathcal{H})}{p(n=0| \text{data}, \mathcal{H})}\\
			&= \frac{\mathcal{Z}_\mathcal{H}}{\mathcal{Z}_\mathcal{H}} \frac{\int \text{d}\bm{\theta}_\mathcal{H} \, p(\text{data} | \bm{\theta}_\mathcal{H}, n=1, \mathcal{H}) \, p(\bm{\theta}_\mathcal{H}, n=1 | \mathcal{H})}{\int \text{d}\bm{\theta}_\mathcal{H} \, p(\text{data} | \bm{\theta}_\mathcal{H}, n=0, \mathcal{H}) \, p(\bm{\theta}_\mathcal{H}, n=0 | \mathcal{H})} \\
			&= \frac{p(n=1 | \mathcal{H})}{p(n=0 | \mathcal{H})} \, \frac{\int \text{d}\bm{\theta}_1\, p(\text{data} | \bm{\theta}_1, \mathcal{H}_1) \, p(\bm{\theta}_1| \mathcal{H}_1)}{\int \text{d}\bm{\theta}_0 \, p(\text{data} | \bm{\theta}_0, \mathcal{H}_0) \, p(\bm{\theta}_0| \mathcal{H}_0)  } \, \frac{\int \text{d} \bm{\theta}_{\overline{1}} \, p(\bm{\theta}_{\overline{1}} | \mathcal{H}_{\overline{1}})}{\int \text{d} \bm{\theta}_{\overline{0}} \, p(\bm{\theta}_{\overline{0}} | \mathcal{H}_{\overline{0}})}\\
			&= \underbrace{\frac{p(n=1 | \mathcal{H})}{p(n=0 | \mathcal{H})}}_{\equiv \, \Pi_{01}} \quad \times \quad  \underbrace{\frac{\mathcal{Z}_1}{\mathcal{Z}_0}}_{\equiv \, \mathcal{B}_{01}} \; .
		\end{align}
		In the last step we used that the inactive parameters, denoted by a bar, do not contribute to the sub-model evidence $\mathcal{Z}_n$. A marginalization over their priors therefore gives one for both sub-models and the last factor is equal to one. As $\mathcal{P}_{01}$ is just the ratio of the number of chain entries after the burn-in period of sub-model 1 compared to sub-model 0 and the model weight ratio $\Pi_{01}$ can be set as a model prior ratio when starting the MCMC chain, the Bayes factor can be obtained by multiplication of the posterior odds ratio with the inverse model weight ratio,
		\begin{align}
			\mathcal{B}_{01} = \mathcal{P}_{01} \times \Pi_{01}^{-1}\; .\label{eq:BF_BF}
		\end{align}
		
\subsection{Implementation details and uncertainties of the computed Bayes factors}

The goal of our calculation is to calculate the Bayes factor as accurately as possible. This can be achieved when the posterior odds ratio $\mathcal{P}_{01}$ is close to one, i.e.~if the hyper-model scan spends about the same amount of time in the two sub-chains. We therefore set the model weight ratio $\Pi_{01}$ to the inverse of the expected Bayes factor and iterate over different weight ratios until the posterior odds ratio is close to one. In practice, this iterative procedure is still an intricate problem due to long runtimes. When the two compared models favour rather distinct regions of the available pulsar-intrinsic red noise parameter space, jumping from one sub-model to the other is initially unlikely. Hence, the burn-in period is long and uncertainties in the computed Bayes factors increase.

Even though our main aim is to calculate Bayes factors for several DSPT scenarios compared to the no common red noise (nCRN) hypothesis, it is faster and more reliable to first compare a given DSPT scenario with the SMBHB hypothesis, see eq.~\eqref{eq:PTA_power_law} ($\gamma  = 13/3$, $A_\text{CP}$ kept as a free parameter). This speeds up the burn-in and results in a more precise computation of the Bayes factor because the posterior distributions for the pulsar-intrinsic red noise parameters are very similar between these two scenarios, unlike when directly comparing to the nCRN hypothesis. As the Bayes factor between a GWB induced by SMBHB and the nCRN hypothesis is known, $\log_{10} \mathcal{B}_\text{SMBHB/nCRN} = 4.5(9)$~\cite{NANOGrav:2020bcs}, and since Bayes factors are multiplicative $\mathcal{B}_{02} = \mathcal{B}_{01} \times \mathcal{B}_{12}$, we can simply rescale our results comparing to SMBHB to the Bayes factor compared to the nCRN hypothesis. In doing so, the burn-in period is shortened by a factor of $\mathcal{O}(10)$. Still, even with a good choice of $\Pi_{01}$ and using this method, the chains take several days to converge to a Bayes factor. Due to these runtime challenges, a publicly available implementation of the described pilot run in~ref.~\cite{PhysRevD.91.044048} to speed up the calculation of Bayes factors would be highly appreciated.

Note however that the method of first comparing to the SMBHB hypothesis does not work if cosmological constraints do not allow a significant GWB in a DSPT model, i.e.~for a stable DS with a large lower boundary of $\beta / H$. Here, the DSPT models favour similar regions in pulsar-intrinsic red noise parameter space as the nCRN hypothesis and a direct comparison would be more advantageous, even though we still compare to the SMBHB model first for consistency. The Bayes factors in these cases therefore have larger uncertainties.

The computation of Bayes factors in itself also comes with statistical uncertainties, in particular from the finite length of the underlying chains. We explicitly checked this uncertainty to be under control by computing the Bayes factor as a function of the number of drawn samples. Using $5 \times 10^6$ samples from the hyper-model (including both sub-models) and conservatively discarding the first $25 \, \%$ due to burn-in, the Bayes factors all converged to a relative uncertainty of a factor of up to $\sim 2$. The convergence rate however depends sensitively on the precise value of the model weights $\Pi_{01}$ as mentioned above. We therefore made sure that the number of samples of both models differed by no more than $\mathcal{O}(10 \, \%)$, which required the model weights $\Pi_{01}$ to be known up to one decimal place.

The continuous lines in figure~\ref{fig:BFs}, depicting the Bayes factors' expected prior dependence, are further prone to uncertainties due to the reduction of points after adapting the prior, cf.~appendix~\ref{sec:priorchoice}. This is particularly relevant when this reduction is large, i.e.~when the Bayes factor is reduced by a significant amount e.g\ for a GWB from bubble wall collisions and a large lower boundary for $\beta / H$. These uncertainties together cause the difference between the expected Bayes factors' prior dependence and the individually calculated Bayes factors for a given prior choice in figure~\ref{fig:BFs}.
		
\subsection{Relating Bayes factors to $p$-values and $Z$-scores}
		\label{sec:z-scores}
		We can express Bayes factors in terms of a $Z$-score to describe the improbability of the null hypothesis~\cite{Athron2020,Fowlie2019}. The $Z$-Score is more commonly known as the ``number of sigmas'' with which some measured quantity deviates from its expectation value. The probability of the null hypothesis can interpreted to be $p(0|\text{data}) = 1 - p(1|\text{data})$ in a frequentist's manner, if one asserts that there is no other possible model to explain the data. If one further interprets the posterior odds ratio to be the ratio of these probabilities $\mathcal{P}_{01} = p(1|\text{data}) / p(0|\text{data})$, the probability of the null hypothesis can equally be expressed as $p(0|\text{data}) = 1/(1 + \mathcal{P}_{01})$. This can be interpreted as a $p$-value, i.e.~the probability to measure data as extreme as the one observed if the null hypothesis were indeed correct. Assuming equal prior probabilities for the two models under comparison, i.e.~$\Pi_{01} = 1$, the $p$-value reads $p = 1/(1 + \mathcal{B}_{01})$. We convert the obtained $p$-value to a $Z$-score using a one-tailed Gaussian,
		\begin{align}
			Z = \Phi^{-1}(1 - p) = \Phi^{-1} \ba{\frac{1}{1 + \frac{1}{\mathcal{B}_{01}}}} \; . \label{eq:z-score}
		\end{align}
		The function $\Phi^{-1}$ is the inverse of the cumulative density function of the standard normal distribution with zero  mean and unit standard deviation. For Bayes factors $\mathcal{B}_{01}$ below 1, implying more evidence for the null hypothesis than for a given more complicated model, we replace $\mathcal{B}_{01}$ by its inverse. We do so to be able to express the tension between the data sets, where the null hypothesis is a better explanation for the observed data than a more complicated model. In that case the $Z$-score should then rather be interpreted as to express the probability of obtaining a signal as low as observed if the signal model were instead correct.
		
	\subsection{Influence of the prior choice on the Bayes factor}
		\label{sec:priorchoice}
		
		The Bayes factor between two models $1$ and $0$ is given by the evidence ratio
		\begin{align}
			\mathcal{B}_{01} = \frac{\mathcal{Z}_1}{\mathcal{Z}_0} \; ,
		\end{align}
		where $\mathcal{Z}_i$ is the evidence of model $i$. We denote the model parameters of model 1 by $\bm{\theta}$, $\mathcal{L}(\bm{\theta})$ is the likelihood to explain the data and $\pi(\bm{\theta})$ is the chosen prior. We here want to investigate the effect of a change in prior $\pi(\bm{\theta}) \rightarrow \tilde{\pi}(\bm{\theta})$ on the Bayes factor $\mathcal{B}_{01} \rightarrow \mathcal{B}_{0{\tilde{1}}}$. Keeping the likelihood and its normalization as well as the number of model parameters unchanged, the Bayes factor differs by a factor
		\begin{align}
			R \equiv \frac{\mathcal{B}_{0{\tilde{1}}}}{\mathcal{B}_{01}} =  \frac{\mathcal{Z}_{\tilde{1}}}{\mathcal{Z}_0}  \cdot \frac{\mathcal{Z}_0}{\mathcal{Z}_1} =  \frac{\int \diff\bm{\theta} \, \mathcal{L}(\bm{\theta}) \,  \tilde{\pi}(\bm{\theta}) }{\int \diff\bm{\theta} \, \mathcal{L}(\bm{\theta}) \,  \pi(\bm{\theta}) } \; .
		\end{align}
		For flat, proper priors (as assumed throughout this work) the effect of a given prior lies in setting the integral boundaries and multiplication with the inverse volume of the model priors $V_\pi$. Using their normalization, the priors can be written as
		\begin{align}
			1 = \int \diff \bm{\theta} \, \pi(\bm{\theta}) = \frac{1}{V_\pi} \int_{V_\pi} \diff \bm{\theta} && \text{and} && 1 = \int \diff \bm{\theta} \, \tilde{\pi} (\bm{\theta}) = \frac{1}{V_{\tilde{\pi}} }   \int_{V_{\tilde{\pi}}} \diff \bm{\theta} \; .
		\end{align}
		In this case the above expression for $R$ simplifies to
		\begin{align}
		R  = \frac{V_\pi}{V_{\tilde{\pi}}} \, \frac{\int_{V_{\tilde{\pi}}} \diff \bm{\theta} \,  \mathcal{L}(\bm{\theta})}{\int_{V_\pi} \diff \bm{\theta} \, \mathcal{L}(\bm{\theta})} \; . \label{eq:BF_prior_dependence1}
		\end{align}
		The effect of changing the range of a flat prior can thus be understood intuitively: Increasing the prior into a region where the likelihood is negligible comes with the cost of increased prior volume, while the posterior integral stays constant. If the likelihood were instead globally flat, an increase in prior volume would have no effect on the Bayes factor, as the increase in the posterior would just be compensated by the increased prior volume. This is just the well-known feature of Bayesian statistics disfavouring unnecessary model complexity. In other words, the cost of introducing a new parameter depends on the coverage of the prior volume with the posterior. If the posterior of this parameter is flat, it can be introduced without changing the Bayes factor. If it however needs to be fine-tuned to fit the data, i.e.~if its posterior is only a thin peak, the coverage of the prior volume is low, reducing the Bayes factor.
		
		The above considerations allow us to reduce the prior ranges after having computed a Bayes factor for some prior ranges that were set too wide, without the need to start a new MCMC chain. This is possible as the chain entries are distributed following the model likelihood $\mathcal{L}(\bm{\theta})$ (being proportional to the posterior within the prior ranges for a flat prior). This means that the ratio of integrals in eq.~\eqref{eq:BF_prior_dependence1} reduces to a ratio of chain entries. We find that the Bayes factor after reducing the prior range is changed by a factor of
		\begin{align}
		R  \rightarrow \frac{V_\pi}{V_{\tilde{\pi}}} \,  \frac{N_{\tilde{\pi}}} {N_\pi} \; ,
		\end{align}
		where $N_{\tilde{\pi}}$ and $N_\pi$ are the number of chain entries enclosed within the prior volumes $V_{\tilde{\pi}}$ and $V_\pi$ respectively. Note that this ``a posteriori'' change of the priors is rather a tool to quickly estimate a small reduction of the prior volume. As soon as one cuts away a region of parameter space that was initially well-covered by the prior, the above approximation comes with an increased statistical error due to potentially low number of samples $N_{\tilde{\pi}}$. This can for instance be seen in the red line in figure~\ref{fig:BFs}. It should also be noted that the uncertainty of this approximation does not scale with $1/\sqrt{N_{\tilde{\pi}}}$ due to the immanent properties of importance sampling which were required to evaluate the ratio of posterior integrals in the first place. A considerable reduction of the prior therefore comes with a large error and still requires the computation of a new MCMC chain with the desired prior ranges.
		
\begin{figure}[t]
			\centering
			\includegraphics[width=.8\linewidth]{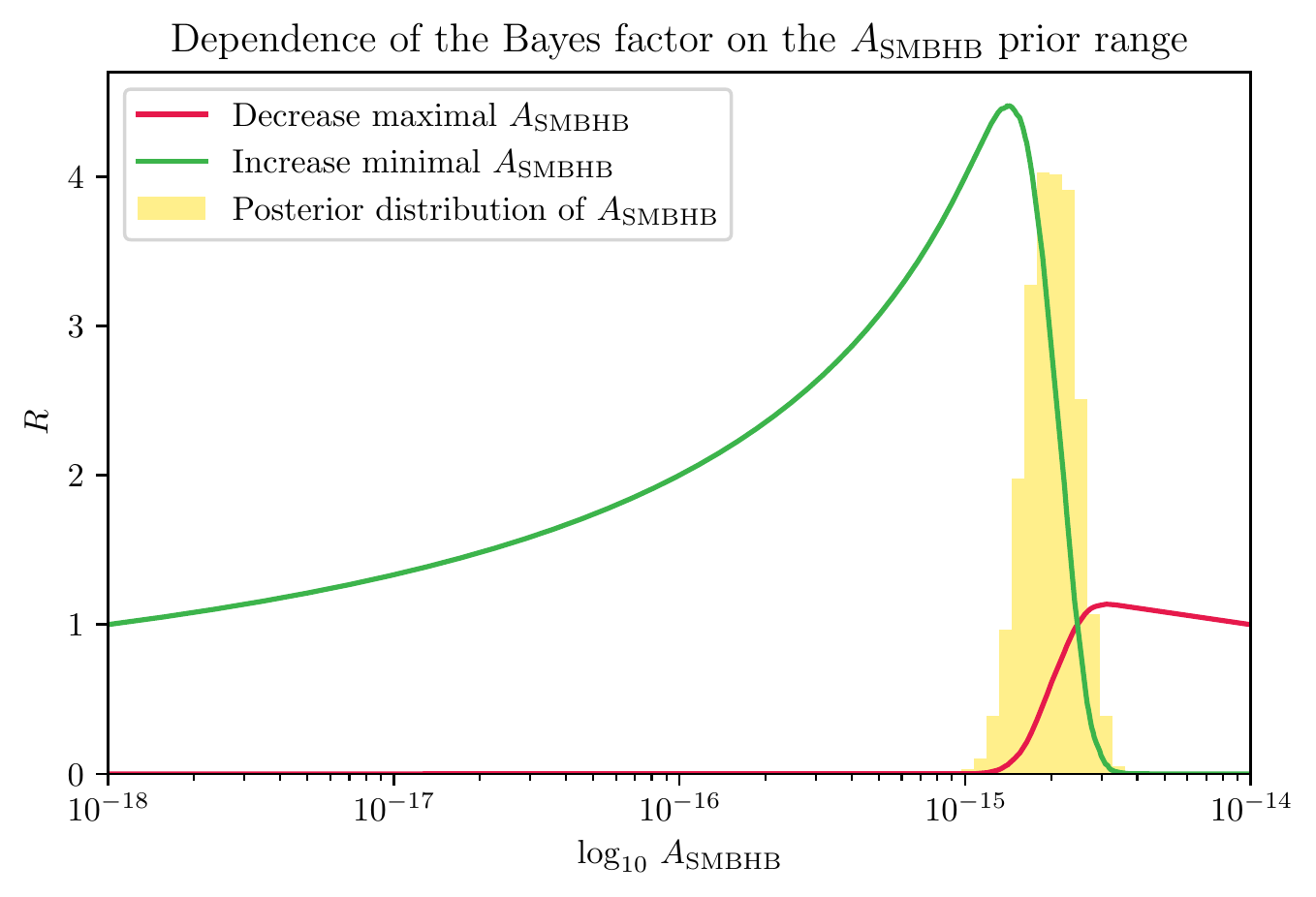}
			\caption{Dependence of the Bayes factor on the $A_\text{SMBHB}$ prior range for a SMBHB signal when increasing the lower boundary (\textcolor{PlotGreen}{green} line) and lowering the upper boundary (\textcolor{PlotRed}{red} line). We also show the posterior distribution of the SMBHB GWB spectral amplitude $A_\text{SMBHB}$ (\textcolor{PlotYellow}{yellow}).}
			\label{fig:priordependence}
		\end{figure}

		To see the effect of the change of the prior range see also figure~\ref{fig:priordependence}, where we show the Bayes factor ratio for SMBHBs with different choices of the lower and upper prior boundaries on the amplitude $A_\text{SMBHB}$. As the posterior for the amplitude peaks between $10^{-15}$ and $10^{-14.5}$, the lower boundary of $10^{-18}$ on the log prior could be doubted to be a good choice, cf.~ref.~\cite{Casey-Clyde2021}. If one instead chose the lower boundary to lie at $10^{-15}$, a factor four increase in the SMBHB's model evidence would be expected. The effect of only lowering the upper prior boundary from $10^{-14}$ down to $10^{-14.5}$ would barely change the Bayes factor due to the little amount of cut-away prior volume not favoured by the posterior distribution.
	
	\subsection{Discussion of the influence of priors on the credibility of a DSPT}
	\label{sec:priorsDSPT}
	
	It is a well-known problem of Bayesian statistics that posteriors and therefore also model evidences and Bayes factors change when a different set of prior probabilities is used. This makes sense in the Bayesian framework, which is intended to update prior beliefs when taking into account measured data. To make sure that our main findings are robust we investigated the influence of the choice of priors on the Bayes factors.
	
	In our main model comparisons that went into producing figure~\ref{fig:BFs}, we employed proper log priors with ranges listed in table~\ref{tab:priors}. All Bayes factors for the decaying DSPT case in figure~\ref{fig:BFs} are below $10^3$. An immediate question arises when comparing to the Bayes factor of $10^{4.5}$ for a GWB from SMBHBs~\cite{NANOGrav:2020bcs} instead of the nCRN hypothesis: Is the discrepancy between the two Bayes factors due to prior-volume effects or due to an actually worse fit to the experimental data by the broken power law with fixed slopes from eq.~\ref{eq:spectra}?
	
	In order to answer this question, we used a twofold approach: First we identified the region of parameter space that results in the highest posterior probabilities. Then we started two chains, a first one in which we constrained all priors to only cover the best-fit regions\footnote{We chose the following log prior ranges in this case: $\alpha_\tot \in \bb{0.02, 0.04}$, $\beta/H \in \bb{1, 3}$, $T_\SM^\perc / \text{MeV} \in \bb{10, 100}$, $\xi_\DS^\perc \in \bb{0.01, 0.1}$, $\tau_\phi / \text{s} \in \bb{0.01, 0.1}$.} and a second one in which we fixed all model parameters to their best-fit values\footnote{Specifically, we chose $\beta/H = 10^{0.3} = 2$, $T_\SM^\perc = 10^{-1.6} \, \text{GeV} = 25 \, \text{MeV}$, $\xi_\DS^\perc = 0.01$, $\tau = 10 \, \text{ms}$. These values were chosen somewhat arbitrarily and are not the result of a precise maximization of the posterior probability. It is conceivable that a slightly better fit to the common red spectrum could be obtained by fine-tuning the parameter points.} except for $\alpha_\tot$, for which we adopted a log prior in the range $\bb{10^{-3}, 10^{1}}$. In doing so the prior in the second approach spans over four orders of magnitude, just as the prior for $A_\text{SMBHB} \in \bb{10^{-18}, 10^{-14}}$, which allows for an easy comparison of the two models on a similar footing.
	
	In the first analysis we obtained a Bayes factor of $\mathcal{O}(5)$ between the decaying DSPT and SMBHB interpretation in favour of the DSPT hypothesis. The second analysis found a Bayes factor of $\mathcal{O}(1.4)$ in favour of the decaying DSPT interpretation. We checked explicitly that this matches the expected Bayes factors from an ``a posteriori'' prior change as described in appendix~\ref{sec:priorchoice}. Both Bayes factors are also consistent with each other, since the one of the second analysis could be enhanced by a factor of $4$ if one decreased the prior range to one decade $\alpha_\tot \in \bb{10^{-2}, 10^{-1}}$, confirming the $\mathcal{O}(5)$ Bayes factor of the first analysis. To make this a fair comparison it should however be noted that also the evidence of the SMBHB hypothesis can be boosted by a relative factor of $4$ by decreasing the prior range of $A_\text{SMBHB}$ from $\bb{10^{-18}, 10^{-14}}$ to $\bb{10^{-15}, 10^{-14}}$, cf.~figure~\ref{fig:priordependence} in appendix~\ref{sec:priorchoice}.
	
	We therefore conclude that a GWB with a broken power-law spectrum with fixed slopes $\Omega_\text{sw}(f) \propto f^3$ and $\Omega_\text{sw}(f) \propto f^{-4}$ for low and high frequencies respectively, can in principle fit the common red spectrum slightly better than a featureless single power-law with spectral index $\gamma_\text{SMBHB} = 13/3$. The origin of the discrepancy between the highest Bayes factors depicted in figure~\ref{fig:BFs} and the reference value of $10^{4.5}$ is due to prior volume effects. These rightfully decrease the credibility of the phase transition explanation, as the required parameter space to explain the common red spectrum is much bigger than that of the SMBHB interpretation.
	
	In particular, the difference in Bayes factors is not due to a worse fit of the pulsar timing data. In fact, in the considered parameter ranges, a decaying DSPT would yield a comparably good (or even slightly better) fit to the timing residuals. If future data is fitted even better by the DSPT explanation compared to SMBHBs, the prior volume effects could be alleviated. With just the NANOGrav 12.5\,yr data set the SMBHB explanation is favoured.

\clearpage

\subsection{Priors for the Bayesian model comparison and calculated Bayes factors}
\label{sec:priors}

\mbox{}
\vspace{12em}
\begin{center}
\begin{sideways}
\begin{minipage}{\linewidth}
		\begin{tabular}{| l | c| c| c| c|c|c|c|c|}
			\hline
			\multicolumn{1}{|c|}{ \textbf{Parameter}} & \multicolumn{3}{|c|}{~~~~~\textbf{Description}~~~~~} & \multicolumn{2}{|c|}{~~~~\textbf{Prior}~~~~} & \multicolumn{3}{|c|}{~~\textbf{Comments}~~} \\
			\hline\hline
			\multicolumn{9}{|c|}{\textbf{PSR-intrinsic red noise}}\\
			\hline\hline
			$A_{\text{red},i}$& \multicolumn{3}{|c|}{Red noise power-law amplitude} & \multicolumn{2}{|c|}{log-Uniform $[-20,-11]$} & \multicolumn{3}{|c|}{one parameter per pulsar} \\
			$\gamma_{\text{red},i}$& \multicolumn{3}{|c|}{Red noise power-law spectral index} & \multicolumn{2}{|c|}{Uniform $[0,7]$} & \multicolumn{3}{|c|}{one parameter per pulsar} \\
			\hline\hline
			\multicolumn{9}{|c|}{\textbf{Supermassive Black Hole Binaries (SMBHBs)}}\\
			\hline\hline
			$A_{\text{SMBHB}}$& \multicolumn{3}{|c|}{Red noise power-law amplitude} & \multicolumn{2}{|c|}{log-Uniform $[-18,-14]$} & \multicolumn{3}{|c|}{one parameter for PTA} \\
			\hline\hline
			\multicolumn{9}{|c|}{\textbf{Stable dark sector phase transition, sound wave or bubble wall collision spectrum}}\\
			\hline\hline
			$\alpha_\text{tot}$& \multicolumn{3}{|c|}{Phase transition strength} & \multicolumn{2}{|c|}{log-Uniform $[-5, 1]$} & \multicolumn{3}{|c|}{one parameter for PTA} \\
			$\beta/H$& \multicolumn{3}{|c|}{Inverse timescale} & \multicolumn{2}{|c|}{
				\begin{tabular}{@{}c@{}}log-Uniform $[0, 3]$, $[\log_{10} 3 , 3]$, \\ $[\log_{10} 5 , 3]$, $[\log_{10} 7, 3]$, $[1, 3]$\end{tabular}			
			} & \multicolumn{3}{|c|}{one parameter for PTA} \\
			$T_\text{SM}^\perc / \text{GeV}$& \multicolumn{3}{|c|}{SM temperature at percolation} & \multicolumn{2}{|c|}{log-Uniform $[-4, 1]$} & \multicolumn{3}{|c|}{one parameter for PTA}\\
			$\xi_\text{DS}^\perc$& \multicolumn{3}{|c|}{DS temperature ratio at percolation} & \multicolumn{2}{|c|}{log-Uniform $[-2, 1]$} & \multicolumn{3}{|c|}{one parameter for PTA}\\
			$g_\text{DS}$& \multicolumn{3}{|c|}{DS degrees of freedom} & \multicolumn{2}{|c|}{Constant $1$} & \multicolumn{3}{|c|}{one parameter for PTA}\\
			$D$& \multicolumn{3}{|c|}{Dilution factor} & \multicolumn{2}{|c|}{Constant $1$} & \multicolumn{3}{|c|}{one parameter for PTA}\\
			\hline\hline
			\multicolumn{9}{|c|}{\textbf{Decaying dark sector phase transition, sound wave spectrum}}\\
			\hline\hline
			$\alpha_\text{tot}$& \multicolumn{3}{|c|}{Phase transition strength} & \multicolumn{2}{|c|}{log-Uniform $[-3, 1]$} & \multicolumn{3}{|c|}{one parameter for PTA} \\
			$\beta/H$& \multicolumn{3}{|c|}{Inverse timescale} & \multicolumn{2}{|c|}{
				\begin{tabular}{@{}c@{}}log-Uniform $[0, 3]$, $[\log_{10} 3 , 3]$, \\ $[\log_{10} 5 , 3]$, $[\log_{10} 7, 3]$, $[1, 3]$\end{tabular}			
		} & \multicolumn{3}{|c|}{one parameter for PTA} \\
			$T_\text{SM}^\perc / \text{GeV}$& \multicolumn{3}{|c|}{SM temperature at percolation} & \multicolumn{2}{|c|}{log-Uniform $[-4, 0]$} & \multicolumn{3}{|c|}{one parameter for PTA}\\
			$\xi_\text{DS}^\perc$& \multicolumn{3}{|c|}{DS temperature ratio at percolation} & \multicolumn{2}{|c|}{log-Uniform $[-3, 1]$} & \multicolumn{3}{|c|}{one parameter for PTA}\\
			$\tau_\phi / \text{s}$& \multicolumn{3}{|c|}{Dark Higgs lifetime} & \multicolumn{2}{|c|}{log-Uniform $[-6,2]$} & \multicolumn{3}{|c|}{one parameter for PTA}\\
			$g_\text{DS}$& \multicolumn{3}{|c|}{DS degrees of freedom} & \multicolumn{2}{|c|}{Constant $1$} & \multicolumn{3}{|c|}{one parameter for PTA}\\
			$D$& \multicolumn{3}{|c|}{Dilution factor} & \multicolumn{2}{|c|}{Constant $1$} & \multicolumn{3}{|c|}{one parameter for PTA}\\
			$m_\phi / \, \text{MeV}$& \multicolumn{3}{|c|}{Dark Higgs mass} & \multicolumn{2}{|c|}{Constant $5$} & \multicolumn{3}{|c|}{one parameter for PTA}\\		
			\hline
	\end{tabular}
	\captionof{table}{Table showing the model parameters together with their respective prior ranges.}
	\label{tab:priors}
\end{minipage}
\end{sideways}
\end{center}

\clearpage

\bibliographystyle{JHEP_improved}
\bibliography{bibliography}

\providecommand{\href}[2]{#2}\begingroup\raggedright\begin{thebibliography}{100}

\bibitem{LIGOScientific:2007fwp}
{\bf LIGO Scientific Collaboration}, B.~P. Abbott et~al.,
  \href{http://dx.doi.org/10.1088/0034-4885/72/7/076901}{{\it {LIGO: The Laser
  interferometer gravitational-wave observatory}}, } {\em Rept. Prog. Phys.}
  {\bf 72} (2009) 076901, [\href{http://arxiv.org/abs/0711.3041}{{\tt
  0711.3041}}].

\bibitem{LISA:2017pwj}
{\bf LISA Collaboration}, P.~Amaro-Seoane et~al., {\it {Laser Interferometer
  Space Antenna}},  \href{http://arxiv.org/abs/1702.00786}{{\tt 1702.00786}}.

\bibitem{NANOGrav:2020bcs}
{\bf NANOGrav Collaboration}, Z.~Arzoumanian et~al.,
  \href{http://dx.doi.org/10.3847/2041-8213/abd401}{{\it {The NANOGrav 12.5 yr
  Data Set: Search for an Isotropic Stochastic Gravitational-wave Background}},
  } {\em Astrophys. J. Lett.} {\bf 905} (2020), no.~2 L34,
  [\href{http://arxiv.org/abs/2009.04496}{{\tt 2009.04496}}].

\bibitem{LIGOScientific:2016aoc}
{\bf LIGO Scientific Collaboration, VIRGO Collaboration}, B.~P. Abbott et~al.,
  \href{http://dx.doi.org/10.1103/PhysRevLett.116.061102}{{\it {Observation of
  Gravitational Waves from a Binary Black Hole Merger}}, } {\em Phys. Rev.
  Lett.} {\bf 116} (2016), no.~6 061102,
  [\href{http://arxiv.org/abs/1602.03837}{{\tt 1602.03837}}].

\bibitem{Burles:2000zk}
S.~Burles, K.~M. Nollett, and M.~S. Turner,
  \href{http://dx.doi.org/10.1086/320251}{{\it {Big bang nucleosynthesis
  predictions for precision cosmology}}, } {\em Astrophys. J. Lett.} {\bf 552}
  (2001) L1--L6, [\href{http://arxiv.org/abs/astro-ph/0010171}{{\tt
  astro-ph/0010171}}].

\bibitem{Auclair2022}
P.~Auclair, D.~Bacon, T.~Baker, T.~Barreiro, N.~Bartolo, et~al., {\it Cosmology
  with the Laser Interferometer Space Antenna},
  \href{http://arxiv.org/abs/2204.05434}{{\tt 2204.05434}}.

\bibitem{Hellings1983}
R.~W. Hellings and G.~S. Downs, \href{http://dx.doi.org/10.1086/183954}{{\it
  {Upper limits on the isotropic gravitational radiation background from pulsar
  timing analysis}}, } {\em Astrophys. J. Lett.} {\bf 265} (1983) L39--L42.

\bibitem{Romano2020}
J.~D. Romano, J.~S. Hazboun, X.~Siemens, and A.~M. Archibald,
  \href{http://dx.doi.org/10.1103/PhysRevD.103.063027}{{\it Common-spectrum
  process versus cross-correlation for gravitational-wave searches using pulsar
  timing arrays}, } \href{http://arxiv.org/abs/2012.03804}{{\tt 2012.03804}}.

\bibitem{Goncharov2021}
B.~Goncharov, R.~M. Shannon, D.~J. Reardon, G.~Hobbs, A.~Zic, et~al.,
  \href{http://dx.doi.org/10.3847/2041-8213/ac17f4}{{\it On the evidence for a
  common-spectrum process in the search for the nanohertz gravitational-wave
  background with the Parkes Pulsar Timing Array}, }
  \href{http://arxiv.org/abs/2107.12112}{{\tt 2107.12112}}.

\bibitem{Chen2021}
S.~Chen, R.~N. Caballero, Y.~J. Guo, A.~Chalumeau, K.~Liu, et~al.,
  \href{http://dx.doi.org/10.1093/mnras/stab2833}{{\it Common-red-signal
  analysis with 24-yr high-precision timing of the European Pulsar Timing
  Array: Inferences in the stochastic gravitational-wave background search}, }
  \href{http://arxiv.org/abs/2110.13184}{{\tt 2110.13184}}.

\bibitem{Antoniadis2022}
J.~Antoniadis, Z.~Arzoumanian, S.~Babak, M.~Bailes, A.~S.~B. Nielsen, et~al.,
  \href{http://dx.doi.org/10.1093/mnras/stab3418}{{\it The International Pulsar
  Timing Array second data release: Search for an isotropic Gravitational Wave
  Background}, } \href{http://arxiv.org/abs/2201.03980}{{\tt 2201.03980}}.

\bibitem{Zic2022}
A.~Zic, G.~Hobbs, R.~M. Shannon, D.~Reardon, B.~Goncharov, et~al., {\it
  Evaluating the prevalence of spurious correlations in pulsar timing array
  datasets},  \href{http://arxiv.org/abs/2207.12237}{{\tt 2207.12237}}.

\bibitem{Pol2020}
N.~S. Pol, S.~R. Taylor, L.~Z. Kelley, S.~J. Vigeland, J.~Simon, et~al.,
  \href{http://dx.doi.org/10.3847/2041-8213/abf2c9}{{\it Astrophysics
  Milestones For Pulsar Timing Array Gravitational Wave Detection}, }
  \href{http://arxiv.org/abs/2010.11950}{{\tt 2010.11950}}.

\bibitem{Casey-Clyde2021}
J.~A. Casey-Clyde, C.~M.~F. Mingarelli, J.~E. Greene, K.~Pardo, M.~Nañez,
  et~al., \href{http://dx.doi.org/10.3847/1538-4357/ac32de}{{\it A quasar-based
  supermassive black hole binary population model: implications for the
  gravitational-wave background}, } \href{http://arxiv.org/abs/2107.11390}{{\tt
  2107.11390}}.

\bibitem{Kelley2016}
L.~Z. Kelley, L.~Blecha, and L.~Hernquist,
  \href{http://dx.doi.org/10.1093/mnras/stw2452}{{\it Massive Black Hole Binary
  Mergers in Dynamical Galactic Environments}, }
  \href{http://arxiv.org/abs/1606.01900}{{\tt 1606.01900}}.

\bibitem{Kelley2017}
L.~Z. Kelley, L.~Blecha, L.~Hernquist, A.~Sesana, and S.~R. Taylor,
  \href{http://dx.doi.org/10.1093/mnras/stx1638}{{\it The Gravitational Wave
  Background from Massive Black Hole Binaries in Illustris: spectral features
  and time to detection with pulsar timing arrays}, }
  \href{http://arxiv.org/abs/1702.02180}{{\tt 1702.02180}}.

\bibitem{Middleton2021}
H.~Middleton, A.~Sesana, S.~Chen, A.~Vecchio, W.~Del~Pozzo, et~al.,
  \href{https://doi.org/10.1093/mnrasl/slab008}{{\it {Massive black hole binary
  systems and the NANOGrav 12.5 yr results}}, } {\em Monthly Notices of the
  Royal Astronomical Society: Letters} {\bf 502} (01, 2021) L99--L103,
  [\href{http://arxiv.org/abs/https://academic.oup.com/mnrasl/article-pdf/502/1/L99/36276026/slab008.pdf}{{\tt
  https://academic.oup.com/mnrasl/article-pdf/502/1/L99/36276026/slab008.pdf}}].

\bibitem{Izquierdo-Villalba2021}
D.~Izquierdo-Villalba, A.~Sesana, S.~Bonoli, and M.~Colpi,
  \href{https://doi.org/10.1093/mnras/stab3239}{{\it {Massive black hole
  evolution models confronting the n-Hz amplitude of the stochastic
  gravitational wave background}}, } {\em Monthly Notices of the Royal
  Astronomical Society} {\bf 509} (11, 2021) 3488--3503,
  [\href{http://arxiv.org/abs/https://academic.oup.com/mnras/article-pdf/509/3/3488/41360112/stab3239.pdf}{{\tt
  https://academic.oup.com/mnras/article-pdf/509/3/3488/41360112/stab3239.pdf}}].

\bibitem{Curylo2021}
M.~Curyło and T.~Bulik,
  \href{http://dx.doi.org/10.1051/0004-6361/202141987}{{\it Predictions for
  LISA and PTA based on SHARK galaxy simulations}, }
  \href{http://arxiv.org/abs/2108.11232}{{\tt 2108.11232}}.

\bibitem{Somalwar2023}
J.~J. Somalwar and V.~Ravi, {\it The origin of the nano-Hertz stochastic
  gravitational wave background: the contribution from $z\gtrsim1$ supermassive
  black-hole binaries},  \href{http://arxiv.org/abs/2306.00898}{{\tt
  2306.00898}}.

\bibitem{Vagnozzi:2020gtf}
S.~Vagnozzi, \href{http://dx.doi.org/10.1093/mnrasl/slaa203}{{\it {Implications
  of the NANOGrav results for inflation}}, } {\em Mon. Not. Roy. Astron. Soc.}
  {\bf 502} (2021), no.~1 L11--L15,
  [\href{http://arxiv.org/abs/2009.13432}{{\tt 2009.13432}}].

\bibitem{Nakai2020}
Y.~Nakai, M.~Suzuki, F.~Takahashi, and M.~Yamada,
  \href{http://dx.doi.org/10.1016/j.physletb.2021.136238}{{\it Gravitational
  Waves and Dark Radiation from Dark Phase Transition: Connecting NANOGrav
  Pulsar Timing Data and Hubble Tension}, }
  \href{http://arxiv.org/abs/2009.09754}{{\tt 2009.09754}}.

\bibitem{Ratzinger:2020koh}
W.~Ratzinger and P.~Schwaller,
  \href{http://dx.doi.org/10.21468/SciPostPhys.10.2.047}{{\it {Whispers from
  the dark side: Confronting light new physics with NANOGrav data}}, } {\em
  SciPost Phys.} {\bf 10} (2021), no.~2 047,
  [\href{http://arxiv.org/abs/2009.11875}{{\tt 2009.11875}}].

\bibitem{NANOGrav:2021flc}
{\bf NANOGrav Collaboration}, Z.~Arzoumanian et~al.,
  \href{http://dx.doi.org/10.1103/PhysRevLett.127.251302}{{\it {Searching for
  Gravitational Waves from Cosmological Phase Transitions with the NANOGrav
  12.5-Year Dataset}}, } {\em Phys. Rev. Lett.} {\bf 127} (2021), no.~25
  251302, [\href{http://arxiv.org/abs/2104.13930}{{\tt 2104.13930}}].

\bibitem{Blasi2020}
S.~Blasi, V.~Brdar, and K.~Schmitz,
  \href{http://dx.doi.org/10.1103/PhysRevLett.126.041305}{{\it Has NANOGrav
  found first evidence for cosmic strings?}, }
  \href{http://arxiv.org/abs/2009.06607}{{\tt 2009.06607}}.

\bibitem{Ellis2020b}
J.~Ellis and M.~Lewicki,
  \href{http://dx.doi.org/10.1103/PhysRevLett.126.041304}{{\it Cosmic String
  Interpretation of NANOGrav Pulsar Timing Data}, }
  \href{http://arxiv.org/abs/2009.06555}{{\tt 2009.06555}}.

\bibitem{Buchmuller:2020lbh}
W.~Buchmuller, V.~Domcke, and K.~Schmitz,
  \href{http://dx.doi.org/10.1016/j.physletb.2020.135914}{{\it {From NANOGrav
  to LIGO with metastable cosmic strings}}, } {\em Phys. Lett. B} {\bf 811}
  (2020) 135914, [\href{http://arxiv.org/abs/2009.10649}{{\tt 2009.10649}}].

\bibitem{Luca2020}
V.~D. Luca, G.~Franciolini, and A.~Riotto,
  \href{http://dx.doi.org/10.1103/PhysRevLett.126.041303}{{\it NANOGrav Hints
  to Primordial Black Holes as Dark Matter}, }
  \href{http://arxiv.org/abs/2009.08268}{{\tt 2009.08268}}.

\bibitem{Kohri2020}
K.~Kohri and T.~Terada,
  \href{http://dx.doi.org/10.1016/j.physletb.2020.136040}{{\it Solar-Mass
  Primordial Black Holes Explain NANOGrav Hint of Gravitational Waves}, }
  \href{http://arxiv.org/abs/2009.11853}{{\tt 2009.11853}}.

\bibitem{Kajantie:1996mn}
K.~Kajantie, M.~Laine, K.~Rummukainen, and M.~E. Shaposhnikov,
  \href{http://dx.doi.org/10.1103/PhysRevLett.77.2887}{{\it {Is there a~ hot
  electroweak phase transition at $m_H \gtrsim m_W$?}}, } {\em Phys. Rev.
  Lett.} {\bf 77} (1996) 2887--2890,
  [\href{http://arxiv.org/abs/hep-ph/9605288}{{\tt hep-ph/9605288}}].

\bibitem{Caprini2015}
C.~Caprini et~al., \href{http://dx.doi.org/10.1088/1475-7516/2016/04/001}{{\it
  {Science with the space-based interferometer eLISA. II: Gravitational waves
  from cosmological phase transitions}}, } {\em JCAP} {\bf 04} (2016) 001,
  [\href{http://arxiv.org/abs/1512.06239}{{\tt 1512.06239}}].

\bibitem{Gori:2022vri}
S.~Gori et~al., {\it {Dark Sector Physics at High-Intensity Experiments}},
  \href{http://arxiv.org/abs/2209.04671}{{\tt 2209.04671}}.

\bibitem{Pospelov:2007mp}
M.~Pospelov, A.~Ritz, and M.~B. Voloshin,
  \href{http://dx.doi.org/10.1016/j.physletb.2008.02.052}{{\it {Secluded WIMP
  Dark Matter}}, } {\em Phys. Lett. B} {\bf 662} (2008) 53--61,
  [\href{http://arxiv.org/abs/0711.4866}{{\tt 0711.4866}}].

\bibitem{Feng:2008mu}
J.~L. Feng, H.~Tu, and H.-B. Yu,
  \href{http://dx.doi.org/10.1088/1475-7516/2008/10/043}{{\it {Thermal Relics
  in Hidden Sectors}}, } {\em JCAP} {\bf 10} (2008) 043,
  [\href{http://arxiv.org/abs/0808.2318}{{\tt 0808.2318}}].

\bibitem{Pospelov:2008zw}
M.~Pospelov, \href{http://dx.doi.org/10.1103/PhysRevD.80.095002}{{\it {Secluded
  U(1) below the weak scale}}, } {\em Phys. Rev. D} {\bf 80} (2009) 095002,
  [\href{http://arxiv.org/abs/0811.1030}{{\tt 0811.1030}}].

\bibitem{Planck2018}
{\bf Planck Collaboration}, N.~Aghanim et~al.,
  \href{http://dx.doi.org/10.1051/0004-6361/201833910}{{\it Planck 2018
  results. VI. Cosmological parameters}, }
  \href{http://arxiv.org/abs/1807.06209}{{\tt 1807.06209}}.

\bibitem{Yeh2022}
T.-H. Yeh, J.~Shelton, K.~A. Olive, and B.~D. Fields,
  \href{http://dx.doi.org/10.1088/1475-7516/2022/10/046}{{\it Probing Physics
  Beyond the Standard Model: Limits from BBN and the CMB Independently and
  Combined}, } \href{http://arxiv.org/abs/2207.13133}{{\tt 2207.13133}}.

\bibitem{Hufnagel:2018bjp}
M.~Hufnagel, K.~Schmidt-Hoberg, and S.~Wild,
  \href{http://dx.doi.org/10.1088/1475-7516/2018/11/032}{{\it {BBN constraints
  on MeV-scale dark sectors. Part II. Electromagnetic decays}}, } {\em JCAP}
  {\bf 11} (2018) 032, [\href{http://arxiv.org/abs/1808.09324}{{\tt
  1808.09324}}].

\bibitem{Forestell:2018txr}
L.~Forestell, D.~E. Morrissey, and G.~White,
  \href{http://dx.doi.org/10.1007/JHEP01(2019)074}{{\it {Limits from BBN on
  Light Electromagnetic Decays}}, } {\em JHEP} {\bf 01} (2019) 074,
  [\href{http://arxiv.org/abs/1809.01179}{{\tt 1809.01179}}].

\bibitem{Depta2020}
P.~F. Depta, M.~Hufnagel, and K.~Schmidt-Hoberg,
  \href{http://dx.doi.org/10.1088/1475-7516/2021/04/011}{{\it Updated BBN
  constraints on electromagnetic decays of MeV-scale particles}, }
  \href{http://arxiv.org/abs/2011.06519}{{\tt 2011.06519}}.

\bibitem{Depta:2020mhj}
P.~F. Depta, M.~Hufnagel, and K.~Schmidt-Hoberg,
  \href{http://dx.doi.org/10.1088/1475-7516/2021/03/061}{{\it {ACROPOLIS: A
  generiC fRamework fOr Photodisintegration Of LIght elementS}}, } {\em JCAP}
  {\bf 03} (2021) 061, [\href{http://arxiv.org/abs/2011.06518}{{\tt
  2011.06518}}].

\bibitem{Kawasaki:2020qxm}
M.~Kawasaki, K.~Kohri, T.~Moroi, K.~Murai, and H.~Murayama,
  \href{http://dx.doi.org/10.1088/1475-7516/2020/12/048}{{\it {Big-bang
  nucleosynthesis with sub-GeV massive decaying particles}}, } {\em JCAP} {\bf
  12} (2020) 048, [\href{http://arxiv.org/abs/2006.14803}{{\tt 2006.14803}}].

\bibitem{Breitbach2019}
M.~Breitbach, J.~Kopp, E.~Madge, T.~Opferkuch, and P.~Schwaller,
  \href{http://dx.doi.org/10.1088/1475-7516/2019/07/007}{{\it {Dark, Cold, and
  Noisy: Constraining Secluded Hidden Sectors with Gravitational Waves}}, }
  {\em JCAP} {\bf 07} (2019) 007, [\href{http://arxiv.org/abs/1811.11175}{{\tt
  1811.11175}}].

\bibitem{Bai2021}
Y.~Bai and M.~Korwar, \href{http://dx.doi.org/10.1103/PhysRevD.105.095015}{{\it
  Cosmological Constraints on First-Order Phase Transitions}, }
  \href{http://arxiv.org/abs/2109.14765}{{\tt 2109.14765}}.

\bibitem{Deng:2023seh}
S.~Deng and L.~Bian, {\it {Constraining low-scale dark phase transitions with
  cosmological observations}},  \href{http://arxiv.org/abs/2304.06576}{{\tt
  2304.06576}}.

\bibitem{Kosowsky1992}
A.~Kosowsky and M.~S. Turner,
  \href{http://dx.doi.org/10.1103/PhysRevD.47.4372}{{\it Gravitational
  Radiation from Colliding Vacuum Bubbles: Envelope Approximation to
  Many-Bubble Collisions}, }
  \href{http://arxiv.org/abs/astro-ph/9211004v1}{{\tt astro-ph/9211004v1}}.

\bibitem{Caprini2019}
C.~Caprini et~al., \href{http://dx.doi.org/10.1088/1475-7516/2020/03/024}{{\it
  {Detecting gravitational waves from cosmological phase transitions with LISA:
  an update}}, } {\em JCAP} {\bf 03} (2020) 024,
  [\href{http://arxiv.org/abs/1910.13125}{{\tt 1910.13125}}].

\bibitem{Huber:2008hg}
S.~J. Huber and T.~Konstandin,
  \href{http://dx.doi.org/10.1088/1475-7516/2008/09/022}{{\it {Gravitational
  Wave Production by Collisions: More Bubbles}}, } {\em JCAP} {\bf 09} (2008)
  022, [\href{http://arxiv.org/abs/0806.1828}{{\tt 0806.1828}}].

\bibitem{Hindmarsh:2013xza}
M.~Hindmarsh, S.~J. Huber, K.~Rummukainen, and D.~J. Weir,
  \href{http://dx.doi.org/10.1103/PhysRevLett.112.041301}{{\it {Gravitational
  waves from the sound of a first order phase transition}}, } {\em Phys. Rev.
  Lett.} {\bf 112} (2014) 041301, [\href{http://arxiv.org/abs/1304.2433}{{\tt
  1304.2433}}].

\bibitem{Hindmarsh:2015qta}
M.~Hindmarsh, S.~J. Huber, K.~Rummukainen, and D.~J. Weir,
  \href{http://dx.doi.org/10.1103/PhysRevD.92.123009}{{\it {Numerical
  simulations of acoustically generated gravitational waves at a first order
  phase transition}}, } {\em Phys. Rev. D} {\bf 92} (2015), no.~12 123009,
  [\href{http://arxiv.org/abs/1504.03291}{{\tt 1504.03291}}].

\bibitem{Hindmarsh2017}
M.~Hindmarsh, S.~J. Huber, K.~Rummukainen, and D.~J. Weir,
  \href{http://dx.doi.org/10.1103/PhysRevD.96.103520
  10.1103/PhysRevD.101.089902}{{\it Shape of the acoustic gravitational wave
  power spectrum from a first order phase transition}, }
  \href{http://arxiv.org/abs/1704.05871}{{\tt 1704.05871}}.

\bibitem{Cutting:2018tjt}
D.~Cutting, M.~Hindmarsh, and D.~J. Weir,
  \href{http://dx.doi.org/10.1103/PhysRevD.97.123513}{{\it {Gravitational waves
  from vacuum first-order phase transitions: from the envelope to the
  lattice}}, } {\em Phys. Rev. D} {\bf 97} (2018), no.~12 123513,
  [\href{http://arxiv.org/abs/1802.05712}{{\tt 1802.05712}}].

\bibitem{Cutting:2019zws}
D.~Cutting, M.~Hindmarsh, and D.~J. Weir,
  \href{http://dx.doi.org/10.1103/PhysRevLett.125.021302}{{\it {Vorticity,
  kinetic energy, and suppressed gravitational wave production in strong first
  order phase transitions}}, } {\em Phys. Rev. Lett.} {\bf 125} (2020), no.~2
  021302, [\href{http://arxiv.org/abs/1906.00480}{{\tt 1906.00480}}].

\bibitem{Jinno:2020eqg}
R.~Jinno, T.~Konstandin, and H.~Rubira,
  \href{http://dx.doi.org/10.1088/1475-7516/2021/04/014}{{\it {A hybrid
  simulation of gravitational wave production in first-order phase
  transitions}}, } {\em JCAP} {\bf 04} (2021) 014,
  [\href{http://arxiv.org/abs/2010.00971}{{\tt 2010.00971}}].

\bibitem{Jinno:2022mie}
R.~Jinno, T.~Konstandin, H.~Rubira, and I.~Stomberg,
  \href{http://dx.doi.org/10.1088/1475-7516/2023/02/011}{{\it {Higgsless
  simulations of cosmological phase transitions and gravitational waves}}, }
  {\em JCAP} {\bf 02} (2023) 011, [\href{http://arxiv.org/abs/2209.04369}{{\tt
  2209.04369}}].

\bibitem{Tenkanen2022}
T.~V.~I. Tenkanen and J.~van~de Vis, {\it Speed of sound in cosmological phase
  transitions and effect on gravitational waves},
  \href{http://arxiv.org/abs/2206.01130}{{\tt 2206.01130}}.

\bibitem{Giese:2020znk}
F.~Giese, T.~Konstandin, K.~Schmitz, and J.~van~de Vis,
  \href{http://dx.doi.org/10.1088/1475-7516/2021/01/072}{{\it
  {Model-independent energy budget for LISA}}, } {\em JCAP} {\bf 01} (2021)
  072, [\href{http://arxiv.org/abs/2010.09744}{{\tt 2010.09744}}].

\bibitem{Giese:2020rtr}
F.~Giese, T.~Konstandin, and J.~van~de Vis,
  \href{http://dx.doi.org/10.1088/1475-7516/2020/07/057}{{\it
  {Model-independent energy budget of cosmological first-order phase
  transitions\textemdash{}A sound argument to go beyond the bag model}}, } {\em
  JCAP} {\bf 07} (2020), no.~07 057,
  [\href{http://arxiv.org/abs/2004.06995}{{\tt 2004.06995}}].

\bibitem{Ertas2021}
F.~Ertas, F.~Kahlhoefer, and C.~Tasillo,
  \href{http://dx.doi.org/10.1088/1475-7516/2022/02/014}{{\it Turn up the
  volume: Listening to phase transitions in hot dark sectors}, }
  \href{http://arxiv.org/abs/2109.06208}{{\tt 2109.06208}}.

\bibitem{Espinosa:2010hh}
J.~R. Espinosa, T.~Konstandin, J.~M. No, and G.~Servant,
  \href{http://dx.doi.org/10.1088/1475-7516/2010/06/028}{{\it {Energy Budget of
  Cosmological First-order Phase Transitions}}, } {\em JCAP} {\bf 06} (2010)
  028, [\href{http://arxiv.org/abs/1004.4187}{{\tt 1004.4187}}].

\bibitem{Bodeker2017}
D.~Bodeker and G.~D. Moore,
  \href{http://dx.doi.org/10.1088/1475-7516/2017/05/025}{{\it {Electroweak
  Bubble Wall Speed Limit}}, } {\em JCAP} {\bf 05} (2017) 025,
  [\href{http://arxiv.org/abs/1703.08215}{{\tt 1703.08215}}].

\bibitem{Bodeker:2009qy}
D.~Bodeker and G.~D. Moore,
  \href{http://dx.doi.org/10.1088/1475-7516/2009/05/009}{{\it {Can electroweak
  bubble walls run away?}}, } {\em JCAP} {\bf 05} (2009) 009,
  [\href{http://arxiv.org/abs/0903.4099}{{\tt 0903.4099}}].

\bibitem{Caprini:2007xq}
C.~Caprini, R.~Durrer, and G.~Servant,
  \href{http://dx.doi.org/10.1103/PhysRevD.77.124015}{{\it {Gravitational wave
  generation from bubble collisions in first-order phase transitions: An
  analytic approach}}, } {\em Phys. Rev. D} {\bf 77} (2008) 124015,
  [\href{http://arxiv.org/abs/0711.2593}{{\tt 0711.2593}}].

\bibitem{Brandenburg:2017neh}
A.~Brandenburg, T.~Kahniashvili, S.~Mandal, A.~Roper~Pol, A.~G. Tevzadze,
  et~al., \href{http://dx.doi.org/10.1103/PhysRevD.96.123528}{{\it {Evolution
  of hydromagnetic turbulence from the electroweak phase transition}}, } {\em
  Phys. Rev. D} {\bf 96} (2017), no.~12 123528,
  [\href{http://arxiv.org/abs/1711.03804}{{\tt 1711.03804}}].

\bibitem{RoperPol:2019wvy}
A.~Roper~Pol, S.~Mandal, A.~Brandenburg, T.~Kahniashvili, and A.~Kosowsky,
  \href{http://dx.doi.org/10.1103/PhysRevD.102.083512}{{\it {Numerical
  simulations of gravitational waves from early-universe turbulence}}, } {\em
  Phys. Rev. D} {\bf 102} (2020), no.~8 083512,
  [\href{http://arxiv.org/abs/1903.08585}{{\tt 1903.08585}}].

\bibitem{Dahl:2021wyk}
J.~Dahl, M.~Hindmarsh, K.~Rummukainen, and D.~J. Weir,
  \href{http://dx.doi.org/10.1103/PhysRevD.106.063511}{{\it {Decay of acoustic
  turbulence in two dimensions and implications for cosmological gravitational
  waves}}, } {\em Phys. Rev. D} {\bf 106} (2022), no.~6 063511,
  [\href{http://arxiv.org/abs/2112.12013}{{\tt 2112.12013}}].

\bibitem{Auclair:2022jod}
P.~Auclair, C.~Caprini, D.~Cutting, M.~Hindmarsh, K.~Rummukainen, et~al.,
  \href{http://dx.doi.org/10.1088/1475-7516/2022/09/029}{{\it {Generation of
  gravitational waves from freely decaying turbulence}}, } {\em JCAP} {\bf 09}
  (2022) 029, [\href{http://arxiv.org/abs/2205.02588}{{\tt 2205.02588}}].

\bibitem{Cirelli2018}
M.~Cirelli, Y.~Gouttenoire, K.~Petraki, and F.~Sala,
  \href{http://dx.doi.org/10.1088/1475-7516/2019/02/014}{{\it {Homeopathic Dark
  Matter, or how diluted heavy substances produce high energy cosmic rays}}, }
  {\em JCAP} {\bf 02} (2019) 014, [\href{http://arxiv.org/abs/1811.03608}{{\tt
  1811.03608}}].

\bibitem{Saikawa2018}
K.~Saikawa and S.~Shirai,
  \href{http://dx.doi.org/10.1088/1475-7516/2018/05/035}{{\it Primordial
  gravitational waves, precisely: The role of thermodynamics in the Standard
  Model}, } \href{http://arxiv.org/abs/1803.01038}{{\tt 1803.01038}}.

\bibitem{Husdal2016}
L.~Husdal, \href{http://dx.doi.org/10.3390/galaxies4040078}{{\it On Effective
  Degrees of Freedom in the Early Universe}, }
  \href{http://arxiv.org/abs/1609.04979}{{\tt 1609.04979}}.

\bibitem{Ellis2020}
J.~Ellis, M.~Lewicki, and J.~M. No,
  \href{http://dx.doi.org/10.1088/1475-7516/2020/07/050}{{\it Gravitational
  waves from first-order cosmological phase transitions: lifetime of the sound
  wave source}, } \href{http://arxiv.org/abs/2003.07360}{{\tt 2003.07360}}.

\bibitem{Breitbach2018}
M.~Breitbach, {\em Gravitational Waves from Cosmological Phase Transitions}.
\newblock mathesis, Johannes Gutenberg University Mainz, Apr., 2018.

\bibitem{Randall:2006py}
L.~Randall and G.~Servant,
  \href{http://dx.doi.org/10.1088/1126-6708/2007/05/054}{{\it {Gravitational
  waves from warped spacetime}}, } {\em JHEP} {\bf 05} (2007) 054,
  [\href{http://arxiv.org/abs/hep-ph/0607158}{{\tt hep-ph/0607158}}].

\bibitem{Konstandin:2011dr}
T.~Konstandin and G.~Servant,
  \href{http://dx.doi.org/10.1088/1475-7516/2011/12/009}{{\it {Cosmological
  Consequences of Nearly Conformal Dynamics at the TeV scale}}, } {\em JCAP}
  {\bf 12} (2011) 009, [\href{http://arxiv.org/abs/1104.4791}{{\tt
  1104.4791}}].

\bibitem{Kierkla2022}
M.~Kierkla, A.~Karam, and B.~Swiezewska,
  \href{http://dx.doi.org/10.1007/JHEP03(2023)007}{{\it Conformal model for
  gravitational waves and dark matter: A status update}, }
  \href{http://arxiv.org/abs/2210.07075}{{\tt 2210.07075}}.

\bibitem{Freese2022}
K.~Freese and M.~W. Winkler,
  \href{http://dx.doi.org/10.1103/PhysRevD.106.103523}{{\it Have Pulsar Timing
  Arrays detected the Hot Big Bang? Gravitational Waves from Strong First Order
  Phase Transitions in the Early Universe}, }
  \href{http://arxiv.org/abs/2208.03330}{{\tt 2208.03330}}.

\bibitem{Hindmarsh:2016lnk}
M.~Hindmarsh, \href{http://dx.doi.org/10.1103/PhysRevLett.120.071301}{{\it
  {Sound shell model for acoustic gravitational wave production at a
  first-order phase transition in the early Universe}}, } {\em Phys. Rev.
  Lett.} {\bf 120} (2018), no.~7 071301,
  [\href{http://arxiv.org/abs/1608.04735}{{\tt 1608.04735}}].

\bibitem{Hindmarsh:2019phv}
M.~Hindmarsh and M.~Hijazi,
  \href{http://dx.doi.org/10.1088/1475-7516/2019/12/062}{{\it {Gravitational
  waves from first order cosmological phase transitions in the Sound Shell
  Model}}, } {\em JCAP} {\bf 12} (2019) 062,
  [\href{http://arxiv.org/abs/1909.10040}{{\tt 1909.10040}}].

\bibitem{Renzini2022}
A.~I. Renzini, B.~Goncharov, A.~C. Jenkins, and P.~M. Meyers, {\it Stochastic
  Gravitational-Wave Backgrounds: Current Detection Efforts and Future
  Prospects},  \href{http://arxiv.org/abs/2202.00178}{{\tt 2202.00178}}.

\bibitem{Phinney2001}
E.~S. Phinney, {\it A Practical Theorem on Gravitational Wave Backgrounds},
  \href{http://arxiv.org/abs/astro-ph/0108028v1}{{\tt astro-ph/0108028v1}}.

\bibitem{Benetti2021}
M.~Benetti, L.~L. Graef, and S.~Vagnozzi,
  \href{http://dx.doi.org/10.1103/PhysRevD.105.043520}{{\it Primordial
  gravitational waves from NANOGrav: a broken power-law approach}, }
  \href{http://arxiv.org/abs/2111.04758}{{\tt 2111.04758}}.

\bibitem{Vaskonen2020}
V.~Vaskonen and H.~Veermäe,
  \href{http://dx.doi.org/10.1103/PhysRevLett.126.051303}{{\it Did NANOGrav see
  a signal from primordial black hole formation?}, }
  \href{http://arxiv.org/abs/2009.07832}{{\tt 2009.07832}}.

\bibitem{Buchmuller2020}
W.~Buchmuller, V.~Domcke, and K.~Schmitz,
  \href{http://dx.doi.org/10.1016/j.physletb.2020.135914}{{\it From NANOGrav to
  LIGO with metastable cosmic strings}, }
  \href{http://arxiv.org/abs/2009.10649}{{\tt 2009.10649}}.

\bibitem{Hindmarsh2020}
M.~B. Hindmarsh, M.~L\"uben, J.~Lumma, and M.~Pauly,
  \href{http://dx.doi.org/10.21468/SciPostPhysLectNotes.24}{{\it {Phase
  transitions in the early universe}}, } {\em SciPost Phys. Lect. Notes} {\bf
  24} (2021) 1, [\href{http://arxiv.org/abs/2008.09136}{{\tt 2008.09136}}].

\bibitem{Wang2022}
D.~Wang, {\it Squeezing Cosmological Phase Transitions with International
  Pulsar Timing Array},  \href{http://arxiv.org/abs/2201.09295}{{\tt
  2201.09295}}.

\bibitem{Wang2022a}
D.~Wang, {\it Novel Physics with International Pulsar Timing Array: Axionlike
  Particles, Domain Walls and Cosmic Strings},
  \href{http://arxiv.org/abs/2203.10959}{{\tt 2203.10959}}.

\bibitem{Ratzinger:2023umd}
W.~Ratzinger, {\em {Misaligned, tilted and distorted: the hard life of audible
  axions}}.
\newblock PhD thesis, Mainz U., 2023.

\bibitem{Dandoy2023}
V.~Dandoy, V.~Domcke, and F.~Rompineve, {\it Search for scalar induced
  gravitational waves in the International Pulsar Timing Array Data Release 2
  and NANOgrav 12.5 years dataset},
  \href{http://arxiv.org/abs/2302.07901}{{\tt 2302.07901}}.

\bibitem{NANOGravTut}
S.~Taylor, S.~Vigeland, J.~Simon, B.~Becsy, and A.~Johnson,
  \href{https://github.com/nanograv/12p5yr_stochastic_analysis}{{\it PTA GWB
  Analysis}, } 2021.

\bibitem{enterprise}
J.~A. Ellis, M.~Vallisneri, S.~R. Taylor, and P.~T. Baker, ``Enterprise:
  Enhanced numerical toolbox enabling a robust pulsar inference suite.''
  Zenodo, Sept., 2020.

\bibitem{enterprise2}
S.~R. Taylor, P.~T. Baker, J.~S. Hazboun, J.~Simon, and S.~J. Vigeland,
  \href{https://github.com/nanograv/enterprise\_extensions}{{\it
  enterprise\_extensions}, } 2021.
\newblock v2.3.3.

\bibitem{justin_ellis_2017_1037579}
J.~Ellis and R.~van Haasteren,
  \href{https://doi.org/10.5281/zenodo.1037579}{{\it PTMCMCSampler: Official
  Release}, } 2017.

\bibitem{harris2020array}
C.~R. Harris, K.~J. Millman, S.~J. van~der Walt, R.~Gommers, P.~Virtanen,
  et~al., \href{https://doi.org/10.1038/s41586-020-2649-2}{{\it Array
  programming with {NumPy}}, } {\em Nature} {\bf 585} (Sept., 2020) 357--362.

\bibitem{2020SciPy-NMeth}
P.~Virtanen, R.~Gommers, T.~E. Oliphant, M.~Haberland, T.~Reddy, et~al.,
  \href{http://dx.doi.org/10.1038/s41592-019-0686-2}{{\it {{SciPy} 1.0:
  Fundamental Algorithms for Scientific Computing in Python}}, } {\em Nature
  Methods} {\bf 17} (2020) 261--272.

\bibitem{Hinton2016}
S.~R. {Hinton}, \href{http://dx.doi.org/10.21105/joss.00045}{{\it
  {ChainConsumer}}, } {\em The Journal of Open Source Software} {\bf 1} (Aug.,
  2016) 00045.

\bibitem{10.1093/mnras/stv2217}
S.~Hee, W.~J. Handley, M.~P. Hobson, and A.~N. Lasenby,
  \href{https://doi.org/10.1093/mnras/stv2217}{{\it {Bayesian model selection
  without evidences: application to the dark energy equation-of-state}}, } {\em
  Monthly Notices of the Royal Astronomical Society} {\bf 455} (11, 2015)
  2461--2473,
  [\href{http://arxiv.org/abs/https://academic.oup.com/mnras/article-pdf/455/3/2461/9377568/stv2217.pdf}{{\tt
  https://academic.oup.com/mnras/article-pdf/455/3/2461/9377568/stv2217.pdf}}].

\bibitem{10.2307/1391010}
S.~J. Godsill, \href{http://www.jstor.org/stable/1391010}{{\it On the
  Relationship between Markov Chain Monte Carlo Methods for Model Uncertainty},
  } {\em Journal of Computational and Graphical Statistics} {\bf 10} (2001),
  no.~2 230--248.

\bibitem{PhysRevD.91.044048}
S.~J. Chamberlin, J.~D.~E. Creighton, X.~Siemens, P.~Demorest, J.~Ellis,
  et~al., \href{https://link.aps.org/doi/10.1103/PhysRevD.91.044048}{{\it
  Time-domain implementation of the optimal cross-correlation statistic for
  stochastic gravitational-wave background searches in pulsar timing data}, }
  {\em Phys. Rev. D} {\bf 91} (Feb, 2015) 044048.

\bibitem{10.2307/2346151}
B.~P. Carlin and S.~Chib, \href{http://www.jstor.org/stable/2346151}{{\it
  Bayesian Model Choice via Markov Chain Monte Carlo Methods}, } {\em Journal
  of the Royal Statistical Society. Series B (Methodological)} {\bf 57} (1995),
  no.~3 473--484.

\bibitem{Fixsen:1996nj}
D.~J. Fixsen, E.~S. Cheng, J.~M. Gales, J.~C. Mather, R.~A. Shafer, et~al.,
  \href{http://dx.doi.org/10.1086/178173}{{\it {The Cosmic Microwave Background
  spectrum from the full COBE FIRAS data set}}, } {\em Astrophys. J.} {\bf 473}
  (1996) 576, [\href{http://arxiv.org/abs/astro-ph/9605054}{{\tt
  astro-ph/9605054}}].

\bibitem{Workman:2022ynf}
{\bf Particle Data Group Collaboration}, R.~L. Workman et~al.,
  \href{http://dx.doi.org/10.1093/ptep/ptac097}{{\it {Review of Particle
  Physics}}, } {\em PTEP} {\bf 2022} (2022) 083C01.

\bibitem{Ramberg2022}
N.~Ramberg, W.~Ratzinger, and P.~Schwaller,
  \href{http://dx.doi.org/10.1088/1475-7516/2023/02/039}{{\it One $\mu$ to rule
  them all: CMB spectral distortions can probe domain walls, cosmic strings and
  low scale phase transitions}, } \href{http://arxiv.org/abs/2209.14313}{{\tt
  2209.14313}}.

\bibitem{Liu2022}
J.~Liu, L.~Bian, R.-G. Cai, Z.-K. Guo, and S.-J. Wang,
  \href{http://dx.doi.org/10.1103/PhysRevLett.130.051001}{{\it Constraining
  first-order phase transitions with curvature perturbations}, }
  \href{http://arxiv.org/abs/2208.14086}{{\tt 2208.14086}}.

\bibitem{Lewicki2023}
M.~Lewicki, P.~Toczek, and V.~Vaskonen, {\it Primordial black holes from strong
  first-order phase transitions},  \href{http://arxiv.org/abs/2305.04924}{{\tt
  2305.04924}}.

\bibitem{Gouttenoire2023}
Y.~Gouttenoire and T.~Volansky, {\it Primordial Black Holes from Supercooled
  Phase Transitions},  \href{http://arxiv.org/abs/2305.04942}{{\tt
  2305.04942}}.

\bibitem{Baker2021}
M.~J. Baker, M.~Breitbach, J.~Kopp, and L.~Mittnacht, {\it Primordial Black
  Holes from First-Order Cosmological Phase Transitions},
  \href{http://arxiv.org/abs/2105.07481}{{\tt 2105.07481}}.

\bibitem{Bennett2020}
J.~J. Bennett, G.~Buldgen, P.~F. de~Salas, M.~Drewes, S.~Gariazzo, et~al.,
  \href{http://dx.doi.org/10.1088/1475-7516/2021/04/073}{{\it Towards a
  precision calculation of $N_{\rm eff}$ in the Standard Model II: Neutrino
  decoupling in the presence of flavour oscillations and finite-temperature
  QED}, } \href{http://arxiv.org/abs/2012.02726}{{\tt 2012.02726}}.

\bibitem{Maggiore:2018sht}
M.~Maggiore, {\em {Gravitational Waves. Vol. 2: Astrophysics and Cosmology}}.
\newblock Oxford University Press, 3, 2018.

\bibitem{Winkler:2018qyg}
M.~W. Winkler, \href{http://dx.doi.org/10.1103/PhysRevD.99.015018}{{\it {Decay
  and detection of a light scalar boson mixing with the Higgs boson}}, } {\em
  Phys. Rev. D} {\bf 99} (2019), no.~1 015018,
  [\href{http://arxiv.org/abs/1809.01876}{{\tt 1809.01876}}].

\bibitem{Ferber:2023iso}
T.~Ferber, A.~Grohsjean, and F.~Kahlhoefer, {\it {Dark Higgs Bosons at
  Colliders}},  \href{http://arxiv.org/abs/2305.16169}{{\tt 2305.16169}}.

\bibitem{Schulze2023}
F.~Schulze, L.~V. Dall'Armi, J.~Lesgourgues, A.~Ricciardone, N.~Bartolo,
  et~al., {\it GW\_CLASS: Cosmological Gravitational Wave Background in the
  Cosmic Linear Anisotropy Solving System},
  \href{http://arxiv.org/abs/2305.01602}{{\tt 2305.01602}}.

\bibitem{Bartolo2022}
N.~Bartolo, D.~Bertacca, R.~Caldwell, C.~R. Contaldi, G.~Cusin, et~al.,
  \href{http://dx.doi.org/10.1088/1475-7516/2022/11/009}{{\it Probing
  Anisotropies of the Stochastic Gravitational Wave Background with LISA}, }
  \href{http://arxiv.org/abs/2201.08782}{{\tt 2201.08782}}.

\bibitem{Taylor2020}
S.~R. Taylor, R.~van Haasteren, and A.~Sesana,
  \href{http://dx.doi.org/10.1103/PhysRevD.102.084039}{{\it From Bright
  Binaries To Bumpy Backgrounds: Mapping Realistic Gravitational Wave Skies
  With Pulsar-Timing Arrays}, } \href{http://arxiv.org/abs/2006.04810}{{\tt
  2006.04810}}.

\bibitem{Kato:2015bye}
R.~Kato and J.~Soda, \href{http://dx.doi.org/10.1103/PhysRevD.93.062003}{{\it
  {Probing circular polarization in stochastic gravitational wave background
  with pulsar timing arrays}}, } {\em Phys. Rev. D} {\bf 93} (2016), no.~6
  062003, [\href{http://arxiv.org/abs/1512.09139}{{\tt 1512.09139}}].

\bibitem{Conneely:2018wis}
C.~Conneely, A.~H. Jaffe, and C.~M.~F. Mingarelli,
  \href{http://dx.doi.org/10.1093/mnras/stz1022}{{\it {On the Amplitude and
  Stokes Parameters of a Stochastic Gravitational-Wave Background}}, } {\em
  Mon. Not. Roy. Astron. Soc.} {\bf 487} (2019), no.~1 562--579,
  [\href{http://arxiv.org/abs/1808.05920}{{\tt 1808.05920}}].

\bibitem{Hotinli:2019tpc}
S.~C. Hotinli, M.~Kamionkowski, and A.~H. Jaffe,
  \href{http://dx.doi.org/10.21105/astro.1904.05348}{{\it {The search for
  anisotropy in the gravitational-wave background with pulsar-timing arrays}},
  } {\em Open J. Astrophys.} {\bf 2} (2019), no.~1 8,
  [\href{http://arxiv.org/abs/1904.05348}{{\tt 1904.05348}}].

\bibitem{Belgacem:2020nda}
E.~Belgacem and M.~Kamionkowski,
  \href{http://dx.doi.org/10.1103/PhysRevD.102.023004}{{\it {Chirality of the
  gravitational-wave background and pulsar-timing arrays}}, } {\em Phys. Rev.
  D} {\bf 102} (2020), no.~2 023004,
  [\href{http://arxiv.org/abs/2004.05480}{{\tt 2004.05480}}].

\bibitem{Sato-Polito:2021efu}
G.~Sato-Polito and M.~Kamionkowski,
  \href{http://dx.doi.org/10.1103/PhysRevD.106.023004}{{\it {Pulsar-timing
  measurement of the circular polarization of the stochastic gravitational-wave
  background}}, } {\em Phys. Rev. D} {\bf 106} (2022), no.~2 023004,
  [\href{http://arxiv.org/abs/2111.05867}{{\tt 2111.05867}}].

\bibitem{ValbusaDallArmi:2023ydl}
L.~Valbusa~Dall'Armi, A.~Nishizawa, A.~Ricciardone, and S.~Matarrese, {\it
  {Circular Polarization of the Astrophysical Gravitational Wave Background}},
  \href{http://arxiv.org/abs/2301.08205}{{\tt 2301.08205}}.

\bibitem{Ellis:2023owy}
J.~Ellis, M.~Fairbairn, G.~H\"utsi, M.~Raidal, J.~Urrutia, et~al., {\it
  {Prospects for Future Binary Black Hole GW Studies in Light of PTA
  Measurements}},  \href{http://arxiv.org/abs/2301.13854}{{\tt 2301.13854}}.

\bibitem{NANOGrav:2023bts}
{\bf NANOGrav Collaboration}, Z.~Arzoumanian et~al., {\it {The NANOGrav
  12.5-year Data Set: Bayesian Limits on Gravitational Waves from Individual
  Supermassive Black Hole Binaries}},
  \href{http://arxiv.org/abs/2301.03608}{{\tt 2301.03608}}.

\bibitem{IPTA:2023ero}
{\bf IPTA Collaboration}, M.~Falxa et~al.,
  \href{http://dx.doi.org/10.1093/mnras/stad812}{{\it {Searching for continuous
  Gravitational Waves in the second data release of the International Pulsar
  Timing Array}}, } {\em Mon. Not. Roy. Astron. Soc.} {\bf 521} (2023), no.~4
  5077--5086, [\href{http://arxiv.org/abs/2303.10767}{{\tt 2303.10767}}].

\bibitem{Abazajian2019}
K.~Abazajian, G.~Addison, P.~Adshead, Z.~Ahmed, S.~W. Allen, et~al., {\it
  CMB-S4 Science Case, Reference Design, and Project Plan},
  \href{http://arxiv.org/abs/1907.04473}{{\tt 1907.04473}}.

\bibitem{Simons2018}
T.~S.~O. Collaboration, P.~Ade, J.~Aguirre, Z.~Ahmed, S.~Aiola, et~al.,
  \href{http://dx.doi.org/10.1088/1475-7516/2019/02/056}{{\it The Simons
  Observatory: Science goals and forecasts}, }
  \href{http://arxiv.org/abs/1808.07445}{{\tt 1808.07445}}.

\bibitem{Dvorkin2022}
C.~Dvorkin, J.~Meyers, P.~Adshead, M.~Amin, C.~A. Argüelles, et~al., {\it The
  Physics of Light Relics},  \href{http://arxiv.org/abs/2203.07943}{{\tt
  2203.07943}}.

\bibitem{Chluba2019}
J.~Chluba, M.~H. Abitbol, N.~Aghanim, Y.~Ali-Haimoud, M.~Alvarez, et~al.,
  \href{http://dx.doi.org/10.1007/s10686-021-09729-5}{{\it New Horizons in
  Cosmology with Spectral Distortions of the Cosmic Microwave Background}, }
  \href{http://arxiv.org/abs/1909.01593}{{\tt 1909.01593}}.

\bibitem{Batell:2009di}
B.~Batell, M.~Pospelov, and A.~Ritz,
  \href{http://dx.doi.org/10.1103/PhysRevD.80.095024}{{\it {Exploring Portals
  to a Hidden Sector Through Fixed Targets}}, } {\em Phys. Rev. D} {\bf 80}
  (2009) 095024, [\href{http://arxiv.org/abs/0906.5614}{{\tt 0906.5614}}].

\bibitem{Heeba2019}
S.~Heeba and F.~Kahlhoefer,
  \href{http://dx.doi.org/10.1103/PhysRevD.101.035043}{{\it Probing the
  freeze-in mechanism in dark matter models with $U(1)^\prime$ gauge
  extensions}, } \href{http://arxiv.org/abs/1908.09834}{{\tt 1908.09834}}.

\bibitem{Bringmann:2020mgx}
T.~Bringmann, P.~F. Depta, M.~Hufnagel, and K.~Schmidt-Hoberg,
  \href{http://dx.doi.org/10.1016/j.physletb.2021.136341}{{\it {Precise dark
  matter relic abundance in decoupled sectors}}, } {\em Phys. Lett. B} {\bf
  817} (2021) 136341, [\href{http://arxiv.org/abs/2007.03696}{{\tt
  2007.03696}}].

\bibitem{Hufnagel2022}
M.~Hufnagel and M.~H.~G. Tytgat, {\it Revisiting the domain of a cannibal dark
  matter},  \href{http://arxiv.org/abs/2212.09759}{{\tt 2212.09759}}.

\bibitem{Athron2020}
P.~Athron, C.~Balázs, A.~Beniwal, J.~E. Camargo-Molina, A.~Fowlie, et~al.,
  \href{http://dx.doi.org/10.1007/JHEP05(2021)159}{{\it Global fits of
  axion-like particles to XENON1T and astrophysical data}, }
  \href{http://arxiv.org/abs/2007.05517}{{\tt 2007.05517}}.

\bibitem{Fowlie2019}
A.~Fowlie, \href{http://dx.doi.org/10.1088/1748-0221/14/10/P10031}{{\it
  Bayesian and frequentist approaches to resonance searches}, }
  \href{http://arxiv.org/abs/1902.03243}{{\tt 1902.03243}}.

\end{thebibliography}\endgroup

\end{document}